\documentclass[manuscript]{acmart}

\setcopyright{none}

\DeclareRobustCommand*\cal{\@fontswitch\relax\mathcal}

\newcommand{\vu}{\mathbf u}

\newcommand{\zero}{\textbf{0}}
\newcommand{\br}{\mathbb{R}}
\newcommand{\BR}{\mathbb{R}}
\newcommand{\conv}{\textbf{conv}}
\newcommand{\st}{\textnormal{s.t.}}

\usepackage[capitalise,noabbrev]{cleveref}

\usepackage[group-separator={,}]{siunitx}
\usepackage{fontawesome}
\usepackage{subfig}
\usepackage{amsfonts}

\renewcommand\cite[1]{\citep{#1}}
\newcommand{\rev}[1]{{\textcolor{black}{#1}}}

\renewcommand{\vv}{\mathbf v}
\renewcommand{\vu}{\mathbf u}

\newcommand{\cI}{{\mathcal{I}}}
\newcommand{\cS}{{\mathcal{S}}}

\newcommand{\R}{{\mathbb{R}}}

\DeclareMathOperator*{\argmax}{arg\,max}

\begin{document}

\title{The Stereotyping Problem in Collaboratively Filtered Recommender Systems}


\author{Wenshuo Guo}
\authornote{Both authors contributed equally to this research.}
\email{wsguo@berkeley.edu}
\author{Karl Krauth}
\authornotemark[1]
\email{karlk@berkeley.edu}
\affiliation{\institution{University of California, Berkeley}
  \state{California}
  \country{USA}
}

\author{Michael I. Jordan}
\affiliation{%
  \institution{University of California, Berkeley}
  \state{California}
  \country{USA}}
\email{jordan@cs.berkeley.edu}

\author{Nikhil Garg}
\affiliation{%
  \institution{Cornell Tech and Technion}
  \state{New York}
  \country{USA}}
\email{ngarg@cornell.edu}


\renewcommand{\shortauthors}{Guo, Krauth, Jordan, and Garg.}

\begin{abstract}
Recommender systems play a crucial role in mediating our access to online information. We show that such algorithms induce a particular kind of stereotyping: if preferences for a \textit{set} of items are anti-correlated in the general user population, then those items may not be recommended together to a user, regardless of that user's preferences and rating history. First, we introduce a notion of \textit{joint accessibility}, which measures the extent to which a set of items can \textit{jointly} be accessed by users. We then study joint accessibility under the standard factorization-based collaborative filtering framework, and provide theoretical necessary and sufficient conditions when joint accessibility is violated. Moreover, we show that these conditions can easily be violated when the users are represented by a \textit{single} feature vector.
To improve joint accessibility,  we further propose an alternative modelling fix, which is designed to capture the diverse multiple interests of each user using a \textit{multi}-vector representation. We conduct extensive experiments on real and simulated datasets, demonstrating the stereotyping problem with standard single-vector matrix factorization models.

\end{abstract}



\maketitle

\section{Introduction}

Recommender systems mediate our access to online information and play a crucial role in online platforms \citep{schafer1999recommender, wei2007survey, zaiane2002building, liu2013soco,hariri2012context,vargas2011effects,levi2012finding}. Especially common are embedding or matrix factorization techniques. In this setting, a recommender system learns a vector representation for each item and user. Then users are recommended the items with feature vectors most similar to their own (e.g., the items that maximize the inner products between the representations)~\citep{takacs2008investigation, koren2009matrix,maneeroj2009hybrid, xue2017deep, zhang2019deep}. The promise of such techniques -- and more broadly collaborative filtering (CF) -- is that we can automatically capture user preferences and item characteristics based on the principle that similar people like similar items. Informally, if Bob and Alice both enjoy item one and Alice further also item two, then Bob is very likely to enjoy item two as well. 


This work is based on a simple insight: collaborative filtering \textit{also} implicitly assumes that negative preferences are transitive. If Bob and Alice both enjoy item one and Alice \textit{dislikes} item two, then a CF approach will learn that Bob is likely to \textit{dislike} item two as well. While such an assumption may seem innocuous, we show that it leads to a \textit{stereotyping} problem. Suppose a user prefers a set of items that are anti-correlated in the broader population. Then these items are unlikely to be jointly recommended to this user -- regardless of that user's preferences and ratings. In other words, for users with multiple interests that are not jointly common in the general population, the recommendation system cannot jointly recommend all the items in which they are interested; the system may instead only recommend them items corresponding to one interest. We note that this problem may especially affect cultural minorities; \textit{if} other users do not like their culture-specific items, then our results imply a recommender system would not simultaneously recommend them both culture-specific items (which are disliked by others) and other items they like (that other users also like).



To formalize this challenge, we first introduce a notion of \textit{joint accessibility}, which indicates when an item set of size $K$ can \textit{jointly} be accessed by a user in a top-$K$ recommendation task (when the $K$ highest predicted items are recommended to the user). Such a notion of joint accessibility captures a fundamental requirement for content diversity in recommender systems: is a user able to access (be recommended) any combination of top $K$ items that they might like? We show that the standard matrix factorization machinery, which models each user and item as a single-vector embedding, does not satisfy joint accessibility conditions. There can exist sets of items that will never be recommended together, even if each item can be individually recommended. In particular, such sets are combinations of items that fall outside the majority of users' preferences. We prove that such a limitation is a consequence of using a single-vector representation for each user -- if two items have representations far apart from one another, then no user vector can be close to both. Under such representations, each user is only able to access a constrained set of item combinations.

We formalize this intuition and analyze its implication theoretically through two geometric interpretations -- providing necessary and sufficient conditions for joint accessibility to hold. Moreover, we show that these conditions are not guaranteed by the standard single-vector representation for users and items used in most CF models. To mitigate this limitation, we provide an alternative multi-vector representation technique in which we represent each user using multiple vectors and prove that it guarantees joint accessibility. This multi-vector representation allows the recommender to learn multiple interests for each user, even if the user's combination of interests is rare.

\textit{Contributions.} 
We introduce the stereotyping problem of \textit{joint accessibility} in recommender systems, and study the theoretical conditions of it and alternative modelling fix. In particular, 

\begin{enumerate}

	\item We formally define the notion of joint accessibility, which can be used as a general measure for auditing recommender systems.
	
	\item We study  joint accessibility under systems using a standard single-vector representation for each user, and provide necessary and sufficient conditions for it to hold. We further show that these conditions can be easily violated with single-vector representations.
	
	\item We then propose an alternative modelling fix. The proposed new modelling technique represents each user with \textit{multiple} feature vectors -- which are learnt to capture user's diverse interests. We analyze the multi-vector representation theoretically, and show that joint accessibility generally holds in such a scheme. 

	\item We conduct extensive experiments on real and simulated datasets, demonstrating the stereotyping problem with standard single-vector matrix factorization models.

\end{enumerate}

\vspace{-5mm}
\subsection*{Related work}\label{sec:related_work}

Recommender systems have been extensively studied, with many successful industrial applications. We discuss the most closely related work under three broad verticals below.

\vspace{-2mm}
\paragraph{Collaborative filtering and matrix factorization.} Collaborative filtering (CF) is one of the most widely used methods for building large-scale recommender systems~\citep{herlocker2004evaluating,schafer2007collaborative}. CF-based recommender systems create recommendations based on what users similar to a given user have liked in the past. At a high level, the similarity metric across users is based on user rating histories, such that the ratings from those like-minded users can be used to predict the ratings of the user of interest; it can also be derived from other items that are likely to be paired with the items which the user of interest has liked in the past. Across many practical implementations of the CF-based approaches, a key tool used for computing the similarities either across the users or items is matrix factorization~\citep{koren2009matrix, xue2017deep,takacs2008investigation, luo2014efficient, yu2012scalable, hernando2016non, wu2018dual}, which provides an efficient way to generate a \textit{single} latent feature vector of each user and item for computing similarities. We illustrate limitations of this approach and in particular the single-vector representation of the users; we then consider an alternative \textit{multi-vector} user representation.

\textit{Content diversity in recommender systems.} A primary goal of traditional recommender systems is to achieve high prediction accuracy. However, recent works have illustrated the pitfalls and limitations of this focus. 
To this end, our work complements a rich line of research on diversity in recommender systems~\citep{fleder2007recommender, cen2020controllable, moller2018not, helberger2018exposure, lathia2010temporal, candillier2011diversity, kunaver2017diversity, di2017adaptive, gravino2019towards, antikacioglu2017post}, that analyze how recommender systems may not show users diverse content and explore alternative methods to improve content diversity. However, most of the prior works assume that each user is represented by single vectors, and consider diversity in the sense of \textit{individual} items. In this paper, we emphasize that such a single-vector embedding can fundamentally limit the ability of the system in capturing a diverse set of interests of each user, and propose a new metric for diversity in terms of accessibility to \textit{sets} of items. We further show how increasing a user's preference for one item could in traditional models decrease the likelihood they receive recommendations for a different item. 

Most related (and the exception to the single-vector assumption) is~\citet{cen2020controllable}, who explore a similar multi-vector user representation framework as us. They train such vectors using modular-based neural networks and provide empirical evidence that such multi-vector representations can better capture the multiple interests for each users. Our paper theoretical uncovers one of the mechanisms of such a solution's benefits, and in the process highlights the heterogeneous effects of single-vector representations on users: users who like sets of items that are together uncommon among other users -- even if all the items individually are common -- normally face especially poor recommendations. 



\textit{Biases in recommender systems.} More broadly, our work is related to the active line of recent works studying biases across different user groups, and long-term impacts such as filter bubbles, glass-ceiling effects, and user agency in recommender systems~\citep{nguyen2014exploring, teppan2015decision, dean2020recommendations, ge2020understanding, mendoza2020evaluating, bellogin2017statistical, jiang2019degenerate, haim2018burst}. The prior works analyze causes of biases theoretically or provide empirical evidence for such biases. Most related is~\citet{dean2020recommendations}, who introduce a measure of reachability and consider user agency in interactive recommender systems; reachability considers whether a \textit{single} item can be recommended to any user, and they study the concept for matrix factorization-based recommender systems through a geometric framework. Our joint accessibility measure generalizes the notion of reachability from single item to combination of multiple items, and thus is able to capture the multiple interests of the users.





\section{Joint Accessibility and its Geometric Interpretation
} \label{sec:single_vec}

In this section, we set up the notations for matrix factorization-based recommendations, and formally define \textit{joint accessibility}. We then present the main theoretical analysis for the single-vector representation of users. In particular, we provide necessary and sufficient conditions for joint accessibility through two geometric interpretations.

\subsection{Basic setup and joint accessibility}

We consider top-$K$ recommendation to a user with a total set of $n$ items, where the user and items are represented by feature vectors with $d$ latent dimensions. In a standard matrix factorization model, each item $j \in [n]$ is associated with a unique feature vector, denoted by $\vv_j \in \BR^d$. The user is also represented by a feature vector $\vu \in \BR^d$.
The system then recommends the items with the $K$ highest predicted ratings based on the feature vectors, which can be learnt from an offline dataset. The user is recommended the items with vectors most similar to their own, i.e. the item that maximizes the inner product in based on the feature representation. Formally, the score (predicted rating) of each item $j$ is the inner product\footnote{More generally, the predicted ratings may further include user and item biases terms.}:
\begin{align}\label{eq:score}  
    s(j) = \vu^\top \vv_j. 
\end{align}
The set of items in a top-$K$ recommendation can thus be denoted as: $S = \argmax_{S \subset [n],\; |S| = K} \sum_{j \in S} s(j).$

With the representation model and its scoring rule at hand, we are now ready to define the notion of \textit{joint accessibility}. In particular, joint accessibility considers whether a set of items can \textit{jointly} be accessed by a user in a recommendation. It captures a fundamental requirement on content diversity in recommender systems: each user -- provided they have shared enough data with the system for it to model their preferences well -- should be able to be recommended any combination of top $K$ items that they possibly like. \rev{Formally, let $\mathcal{I}$ be the set of all items, i.e. $\mathcal{I} = \{\vv_j\}_{j=1}^n$.  Let $\cI_K(\{\vv_j\}_{j=1}^n)$ be the space of all subsets of items with size $K$, i.e., $\cI_K(\{\vv_j\}_{j=1}^n) \triangleq \{S: S \subseteq \cI,\; |S| = K\}$.} We define joint accessibility for a top-$K$ recommendation to a user as follows. 


\begin{definition}(\textit{Joint accessibility})\label{def:joint-access} Consider a top-$K$ recommendation to a user with a total of $n$ items. 
Then a recommender system with item vectors $\{\vv_j\}_{j=1}^n$ satisfies joint accessibility if and only if
\[ \forall S\in \cI_K(\{\vv_j\}_{j=1}^n),\; \text{ it is true that } \exists \vu \in \BR^d\;
\text{ s.t. } S \text{ is recommended.}
\]

We further say that a \textit{given} set $S$ is jointly accessible if there exists $ \vu \in \BR^d\;
\text{ s.t. } S \text{ is recommended.}$
\end{definition}

Intuitively, joint accessibility of a top-$K$ recommendation means that for any item sets with size $K$, there exists at least one user vector for which
this sets of items will be recommended. In other words, if joint accessibility of a top-$K$ recommendation is not satisfied, then there exist some sets of items with size $K$, such that these sets of items will never be chosen jointly by the recommender system -- even if there exists a user whose history suggests they like exactly that combination of items. In the coming subsections, we formally characterize the necessary and sufficient conditions for joint accessibility to be satisfied, through two geometric interpretations.




\subsection{Accessibility conditions and their geometric interpretations}\label{sec:geometric}

Definition~\ref{def:joint-access} defines the joint accessibility of a diverse set of items through the existence of a feasible user vector. In the next two subsections, we show that such an existence can be conveniently translated through two geometric interpretations to properties that are easier to visualize. These geometric interpretations provide necessary and sufficient conditions for joint accessibility for a given set of items. They further illustrate why dissimilar items in particular may not be recommended together, regardless of a user's interests. 

\textit{Convex hull interpretation.} Our first geometric interpretation uses an argument with convex hulls of vectors. This geometric interpretation extends the analysis in~\citet{dean2020recommendations} to \textit{joint} accessibility of multiple items.

\begin{figure*}[!ht]
\centering
\begin{tabular}{cc} 
\includegraphics[width=0.28\textwidth]{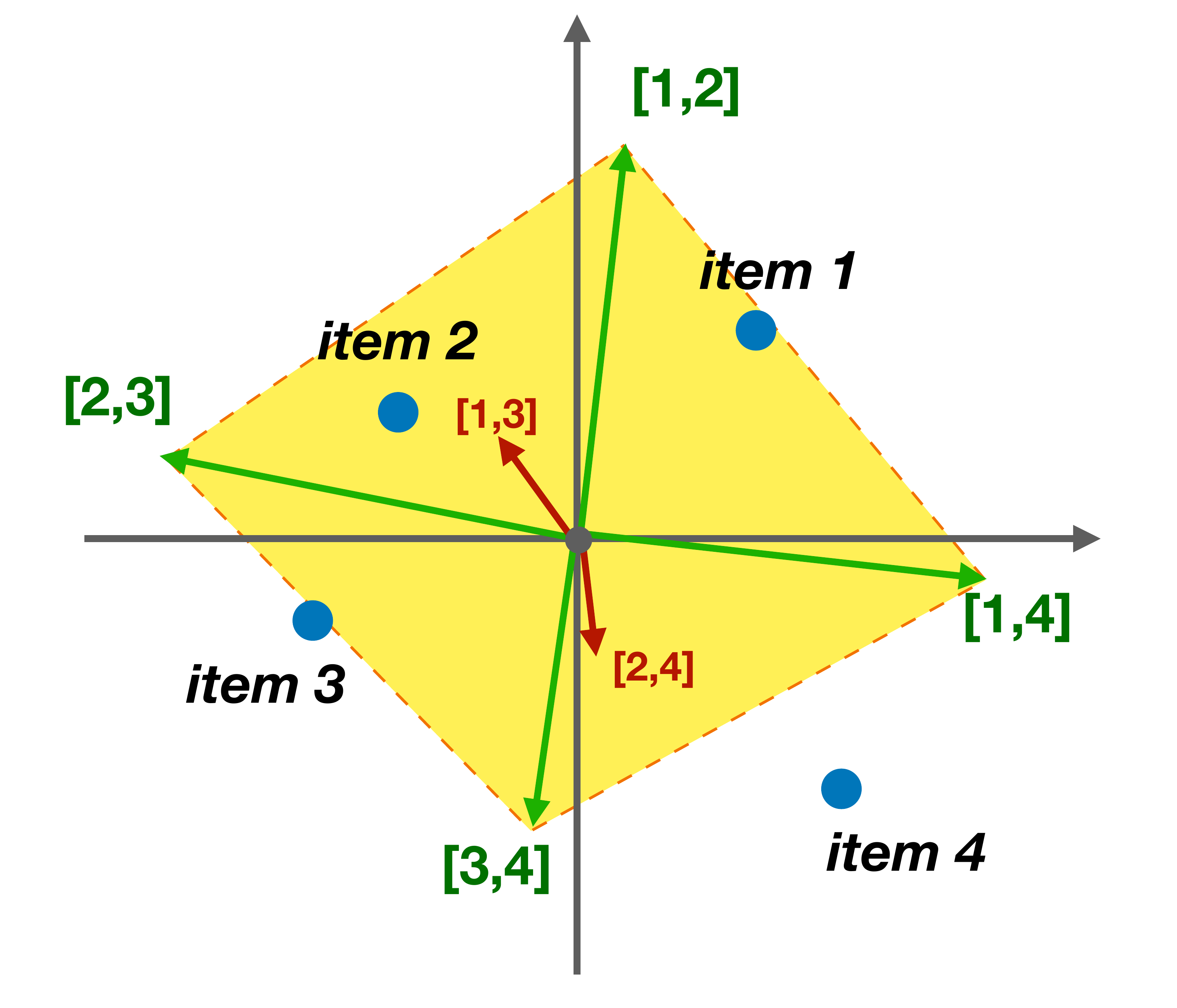} &
\includegraphics[width=0.33\textwidth]{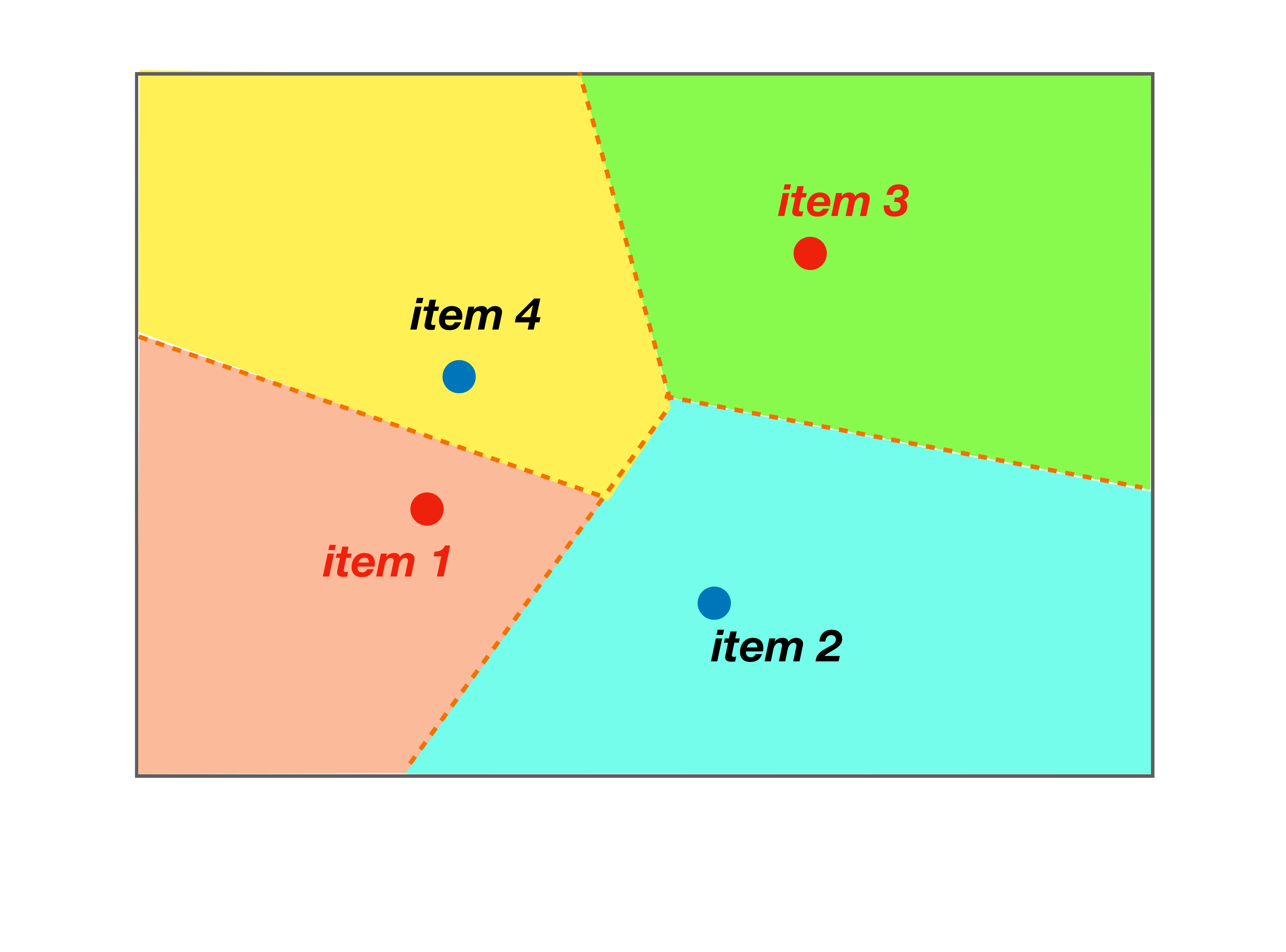} 
\end{tabular}
\caption{Geometric interpretations of joint accessibility for Top-$2$ recommendations. \textit{(left)} convex hull interpretation. Green vectors represent sum vectors for item pairs.; \textit{(right)} Voronoi diagram interpretation. Colored blocks represent Voronoi cells. }\label{fig:convexhull-voronoi} 

\Description{The left figure shows the convex hull of four item vectors in two dimension. The four item vectors each resides in a quadrant of the graph. However, the sum of vector one and vector three, and the sum of vector two and four both are very close to the origin and are not vertices of the convex hull. This indicates that these two item pairs are not jointly accessible. The right figure shows the Voronoi diagram of four item vectors in two dimension. Each item lies in its Voronoi cell, and the cell of item one and item three are not neighbors. This indicates that item one and item three are not jointly accessible in this case.}

\vspace{-5mm}
\end{figure*}

\begin{theorem}(Proof in Appendix~\ref{app:proof:thm:single-vec-vertex-condition})\label{prop:single-vec-vertex-condition} (Single vector accessibility condition) Given $n$ items with feature vectors $\vv_1, \cdots \vv_n \in \br^d$, an item set $S$ of size $K$ is accessible if and only if $\sum_{i \in S} \vv_i$ is a vertex of the convex hull: $\conv\{\sum_{i \in S} \vv_i, \forall S \subseteq [n], |S| = K\}$.
\end{theorem}

Theorem~\ref{prop:single-vec-vertex-condition} translates accessibility of a certain set of items to a linear constraint in the items' feature vectors. Intuitively, $\conv\{\sum_{i \in S} \vv_i, \forall S \subseteq [n], |S| = K\}$ is the convex hull of all the possible sums of $K$ item vectors.  Theorem~\ref{prop:single-vec-vertex-condition} guarantees that, if the sum of the feature vectors of $K$ items is not a vertex of this convex hull, this set of items can not be accessed to any user. As illustrated in Figure~\ref{fig:convexhull-voronoi} \textit{(left)}, with top-2 recommendation with latent dimension $d=2$, the pair of item 1 and item 3, and the pair of item 2 and item 4 fall inside the convex hull. They are are thus not jointly accessible in this configuration. Note that this joint inaccessibility holds even though every \textit{individual} item can be recommended to users: each of the four items are on the convex hull defined by their individual item vectors.  

\textit{Voronoi diagram interpretation.} The convex hull of a general set of vectors can be difficult to visualize. For instance, for top-2 recommendation as illustrated in  Figure~\ref{fig:convexhull-voronoi} \textit{(left)}, we would need to visualize the sum of $O(n^2)$ pairs of item vectors and the corresponding convex hull. Therefore, we accompany the convex hull interpretation with a second geometric interpretation, using Voronoi diagrams. This interpretation is more easier to visualize especially when $K$ is small, though holds in a more limited setting. Specifically, we illustrate the alternative geometric interpretation for top-2 recommendation. Let us consider item feature vectors in $\R^d$ and a recommendation algorithm with the scoring rule as in Eq~\eqref{eq:score}. We further assume that the feature vectors are normalized and thus locate on a sphere in $\R^d$. These item vectors on the sphere then define a Voronoi diagram, as follows. 

\begin{definition} (Spherical Voronoi diagram for item vectors, \citep{de1997computational})
Denote a unit sphere in $\R^d$ as $S^d$. A spherical Voronoi diagram for item vectors is defined with $n$ sites located at the $n$ item vectors, and with the distance on the surface of the sphere. The Voronoi cell associated with the item vector $\vv_k$ is the set of all points on the sphere whose distance to $\vv_k$ is not greater than their distance to the any other sites $j$, where $j \neq k$. Two items are Voronoi neighbors if they have neighboring Voronoi cells.
\end{definition}


\begin{theorem}\label{thm:voronoi}(Proof in Appendix~\ref{app:proof:thm:voronoi})
Given $n$ items with normalized feature vectors $\vv_1, \cdots \vv_n \in \br^d$, an item set $S$ of size $K=2$ is accessible if and only if the items are Voronoi neighbors.
\end{theorem}

Theorem~\ref{thm:voronoi} provides another interpretation for joint accessibility of item pairs. To see this, Figure~\ref{fig:convexhull-voronoi} \textit{(right)} illustrates the Voronoi cells for the item vectors in $\R^3$ by showing the two-dimensional surface of the sphere. We see that item 1 and item 3 are not Voronoi neighbors, and therefore not jointly accessible. Intuitively, if two item vectors are far apart, then no single user vector can be simultaneously close to both item vectors. That two items are Voronoi neighbors establishes that there exist user vectors at their joint boundary whose closest neighbors to the user vector are those two items. Otherwise, if items are not Voronoi neighbors, no such user vector exists.

We further provide an observation on the transitiveness of the impossibility for joint accessibility. Intuitively, if joint accessibility is violated for a smaller set of items, it is also violated if we have more items available to recommend. For top-$K$ recommendation with a total of $m$ items, if joint accessibility is violated for a pair of values of $n,d$, then it is also violated for $d$ and any $n'\geq n$. Therefore, guaranteeing joint accessibility with a small set of item is essential for joint accessibility of a large set of available recommendations. 

\section{Accessibility with Single and Multi-vector Representations}
\label{sec:multi_vec}

In the previous section, we considered accessibility of a specific subset of items. In this section, we consider joint accessibility across all subsets. We first do so for the standard matrix factorization framework where each user is represented by a single vector, formalizing the impossibility examples from the previous section. Then, we propose an alternative representation model for mitigating the stereotype problem in joint accessibility. This new model represents each user with multiple vectors in particular can capture each user's diverse multiple interests. We then analyze the multi-vector representation theoretically, providing sufficient conditions for joint accessibility. Note that as pointed out by \citet{dean2020recommendations}, accessibility to a single item is not always guaranteed, if the vector is not on the convex hull defined by the individual item vectors. In this section, we thus ask: \textit{(when) does accessibility of all individual items imply that all subsets of items are jointly accessible?} 

\subsection{Impossibility result with single-vector representation}

Our first result formalizes the examples from the previous section: that under a standard matrix factorization framework, in general individual accessibility does not imply joint accessibility. 

\begin{theorem}\label{thm:single-vec-impossibility}(Proof in Appendix~\ref{app:proof:thm:single-vec-impossibility})
Consider a single-vector user representation, with $n$ items, and top-K recommendation. Then, there exists item feature vectors $\vv_1, \cdots \vv_n \in \br^d$ such that every item is individually accessible, but that there exists a set of $K$ items that is not jointly accessible. 
\end{theorem}

Theorem~\ref{thm:single-vec-impossibility} shows a fundamental limitation of the single item representation: \rev{it is possible that -- regardless of a user's individual history and thus user vector $u$ -- a set of items that the user likes cannot be jointly recommended to the user. In our empirical analysis in \Cref{sec:exp}, we show that this setting is common: a large proportion of individually accessible items cannot be jointly accessed (even in pairs), \textit{regardless} of a user's history.} 
\rev{We view such an occurrence as a type of \textit{stereotyping}: if two items $i$ and $j$ are anti-correlated in the general population and thus not Voronoi neighbors in the learned representation, the system imposes that those who like item $i$ \textit{cannot} like item $j$, even if an individual's personal ratings history suggests otherwise. }

\rev{As previewed in our discussion above and formally shown in the empirical analysis below, we note that this situation may especially affect cultural minorities. Suppose most members of the majority group do not like certain items from the minority culture. Then, in the learned feature representation, the item vectors for generally liked items or items unrelated to either culture are likely to be far from the vectors for items form the minority culture -- by design, matrix factorization methods learn such relationships for lower prediction error. In such cases, individuals who express an interest in items from the minority culture  \textit{cannot} be recommended other items, even if their individual ratings histories (or even the histories of other cultural minorities) would indicate that they like such items. In this manner -- due to the behavior of the majority group -- the algorithm stereotypes members of the minority group as only liking a subset of items not liked by the majority group.}

\subsection{Multi-vector representation}

From the geometric interpretation analysis in Section~\ref{sec:geometric}, we see that the limitation of joint accessibility depends crucially on the single-vector representation: the same user vector cannot be simultaneously close to two far-apart vectors, and so such a representation is restricted in capturing multiple interests of the users. Therefore, our new framework uses multiple vectors to represent each user.
Consider a top-$K$ recommendation to a user with a total of $n$ items, where the user and item vectors have latent dimensions $d$. In a multi-vector representation model, we associate a user with $m$ feature vectors, $\{\vu_i, \ldots, \vu_m\} \in \BR^d$.
The system then recommends the top $K$ items based on the feature vectors. An important change is that the new scoring rule takes the \textit{maximum} predicted rating over all user vectors for an item as the final prediction; thus a low score in one user vector does not matter. Formally, the score of each item $j$ is
\begin{align}\label{eq:score-multi}  
    s(j) =\max_{i\in [m]} \vu_i^\top \vv_j.
\end{align}
The set of items in an top $K$ recommendation are those with the $K$ highest scores: $S \triangleq \argmax_{S \subset [n],\; |S| = K} \sum_{j \in S} s(j)$. With such a representation and with $m \geq K$, individual accessibility guarantees joint accessibility. 

\begin{theorem}(Proof in Appendix~\ref{app:proof:thm:multi-vec-access})\label{thm:multi-vec-access}
Consider $m$-vector user representations and the recommender system scoring rule as in Eq~\eqref{eq:score-multi}, with $n$ items, and top-K recommendation. Suppose $m \geq K$. Then, for any item feature representations $\vv_1, \cdots \vv_n \in \br^d$, if every item is individually accessible, then every set of size $K$ items is jointly accessible.
\end{theorem}

The result follows immediately from the definitions. If there exist individual user vectors such that each item is recommended, then for any set of size $K$ there exists a set of user vectors of size $K$ such that the set is recommended. 








\section{Experiments}\label{sec:exp}


We empirically show that standard matrix factorization techniques lead to joint accessibility issues. In \Cref{sec:empmethods}, we outline our empirical setting. We include the results in \Cref{sec:empresults_inaccesibility}. 

\subsection{Empirical setting and methods}
\label{sec:empmethods}

We first illustrate the insights and results using a synthetically generated dataset, and then using MovieLens 10M dataset (ML10M)~\citep{harper2015movielens}. The synthetic setting allows us to observe ground truth preferences, and thus to illustrate exactly which items are not jointly accessible and in turn which users receive insufficiently diverse recommendations. The ML10M setting further confirms that our insights extend to real recommendation settings. 

\paragraph{Data characteristics.} Here we briefly describe our datasets and training procedures. More details are in \Cref{app:exp-additional-details}. \rev{The MovieLens 10M dataset (ML10M) contains 10,000,054 ratings and 95580 tags applied to 10681 movies by 71567 users of the online movie recommender service \textit{MovieLens} . As a standard benchmark, we used the edx dataset (train set) of it, which contains 9,000,055 ratings, with of 10,677 movies by 69,878 users, and consisted of 90\% of the original benchmark MovieLens 10M dataset. In particular, the users were selected at random for inclusion, and all users selected had rated at least 20 movies. 
}

\rev{We further generate two synthetic datasets using a modified version of the \texttt{latent-static} environment from the RecLab simulation platform \citep{krauth2020offline}. For the first, we sample \num{10677} items and  \num{69878} users, matching the numbers from ML10M; for the second, we sample \num{500} items and  \num{20000} users. As the ground truth, we represent each item as a \num{64} dimensional vector. To illustrate exactly how single-vector representation fails when users might have different (potentially ``opposing'' interests), we construct for each user a ground truth representation as two \num{64} dimensional vectors, as detailed below. Then, we sample ratings (10\% of all item-user pairs) by adding noise to the ground truth rating according to these representations (where, as in our user model, the user's rating for an item is the max across the ratings implied by either of their representations).}

\rev{The purpose of the first synthetic dataset is to understand the general limitations of single-vector representations. As such, we simply draw all vectors (both vectors for each user, and each item vector) uniformly at random from the unit ball.} 
\rev{The purpose of the second synthetic dataset is to specifically understand the differential effect on minority users and to demonstrate the stereotyping effect discussed above. In this case, we first draw five topic vectors, where one of the topic vectors (the minority affiliated topic) is designed to have low similarity with the other topics. Then, we regard 20\% of users as minority users. Each majority user has ground truth vectors drawn from noise around two of the majority topic vectors. The minority users each have one vector drawn from noise around the minority topic vector, the other from around one of the majority topic vectors. Note that, by design, minority users also rate highly items from one of the majority topics, but are the only ones to also rate highly items from the minority topic.}

We train matrix factorization models using the standard alternating least squares (ALS) algorithm \citep{hu2008collaborative}. \rev{After obtaining the trained user and item vectors, for each dataset we select a subset of the \num{400} most popular individual items for which to carry out our analysis (for computational reasons, as our metrics calculations are quadratic in the number of items; analysis with 400 items chosen uniformly at random are in the Appendix and are qualitatively similar)}. For this subset and all users, we calculate a predicted rating between that user $i$ and item $j$, as the dot product of the given item and user vectors, $u_i \cdot v_j$. Finally, we find the pair of items that would be recommended to each user with top-2 recommendations, i.e., the two items with the highest predicted rating for that user. For the synthetic datasets, we also find the pair of items that an oracle recommender would recommend, i.e., the top two items in the user's true ranking over the items.  We include the complete data generation details and training details in Appendix~\ref{app:exp-additional-details}.

\subsection{Analysis and results}
\label{sec:empresults_inaccesibility}

The goal of our experimental analysis is to illustrate our insights regarding joint accessibility and the stereotyping challenge: first, that sets of items are not (and cannot be) jointly recommended together even if every individual item can be recommended; second, that in particular the sets that are inaccessible are those that in the training data are anti-correlated, and so recommendations do not reflect the underlying diversity of a given user's preferences; \rev{and third, that cultural minorities in particular are stereotyped into one topic.} Note that, by construction of the synthetic datasets, if recommendations were made according to the oracle, there would not be a strong relationship between how similar two item vectors are and how often they're recommended together; an optimal multiple vector model could recommend dissimilar items together. In contrast, in all the datasets using the trained single-representation vectors, we indeed observe that only items that are similar are in practice recommended together. We illustrate this in several ways, starting with the synthetic settings where we can also compare to what an oracle recommender would do.

\paragraph{First synthetic setting.}
\begin{figure}[!ht]
 \vspace{-3mm}
	\centering
	\subfloat[Subfigure 1 list of figures text][Number of times for pairs of items with a given similarity are recommended together. For example, items $j,k$ with true similarity $\vv_j \cdot \vv_k = 1$ are on average recommended together for 1 user in the training set, using either the oracle or the trained recommendations. ]{
		\includegraphics[height=0.19\linewidth]{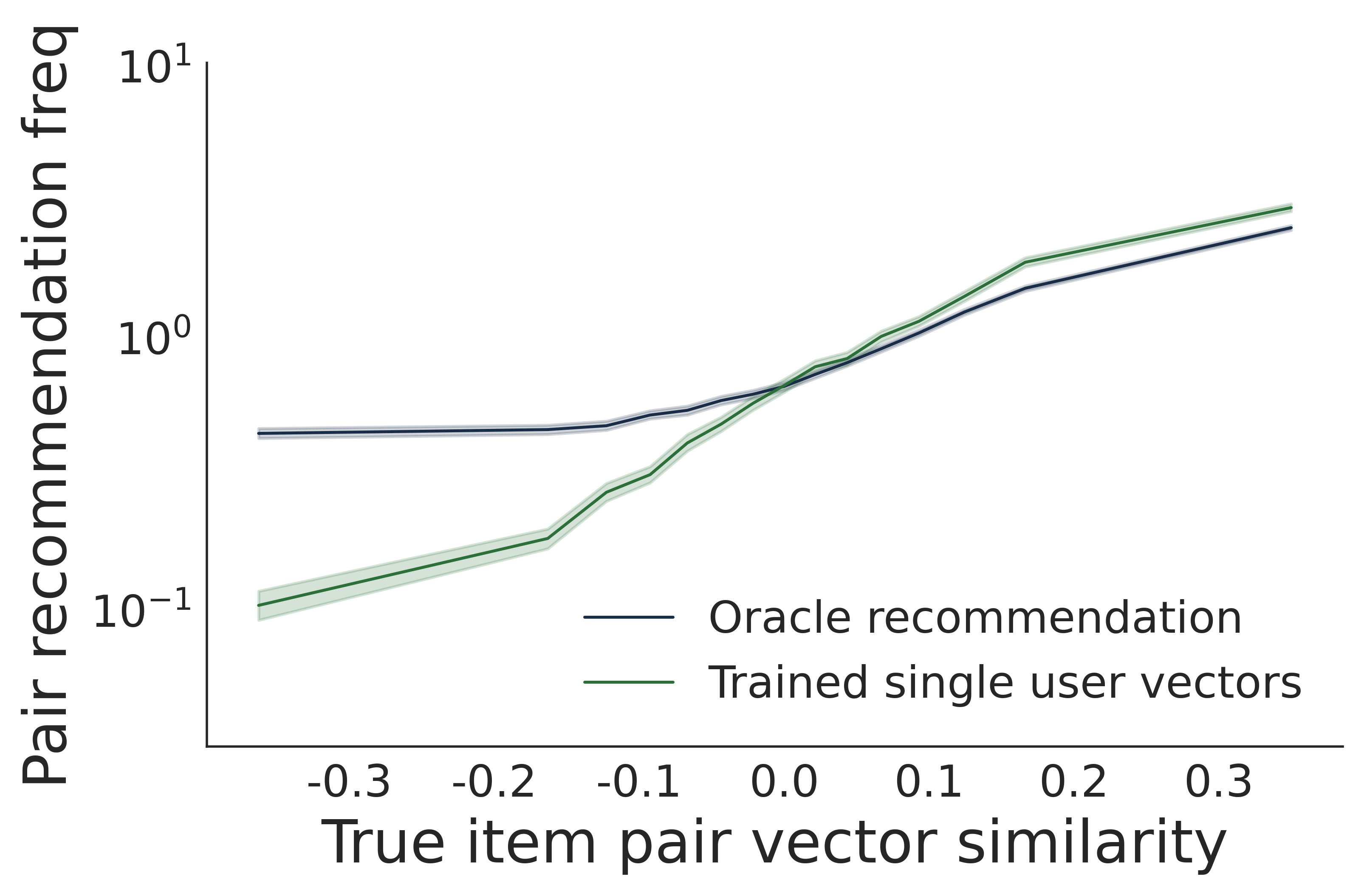}
		\label{fig:empiricalappearanceprob}}
	\qquad
	\subfloat[Subfigure 2 list of figures text][\rev{Fraction of} item pairs with a given similarity \rev{that} are recommended together for a user vector constructed specifically to maximize the ratings for that pair. With a single vector per user, not all items are accessible.]{
		\includegraphics[height=0.19\linewidth]{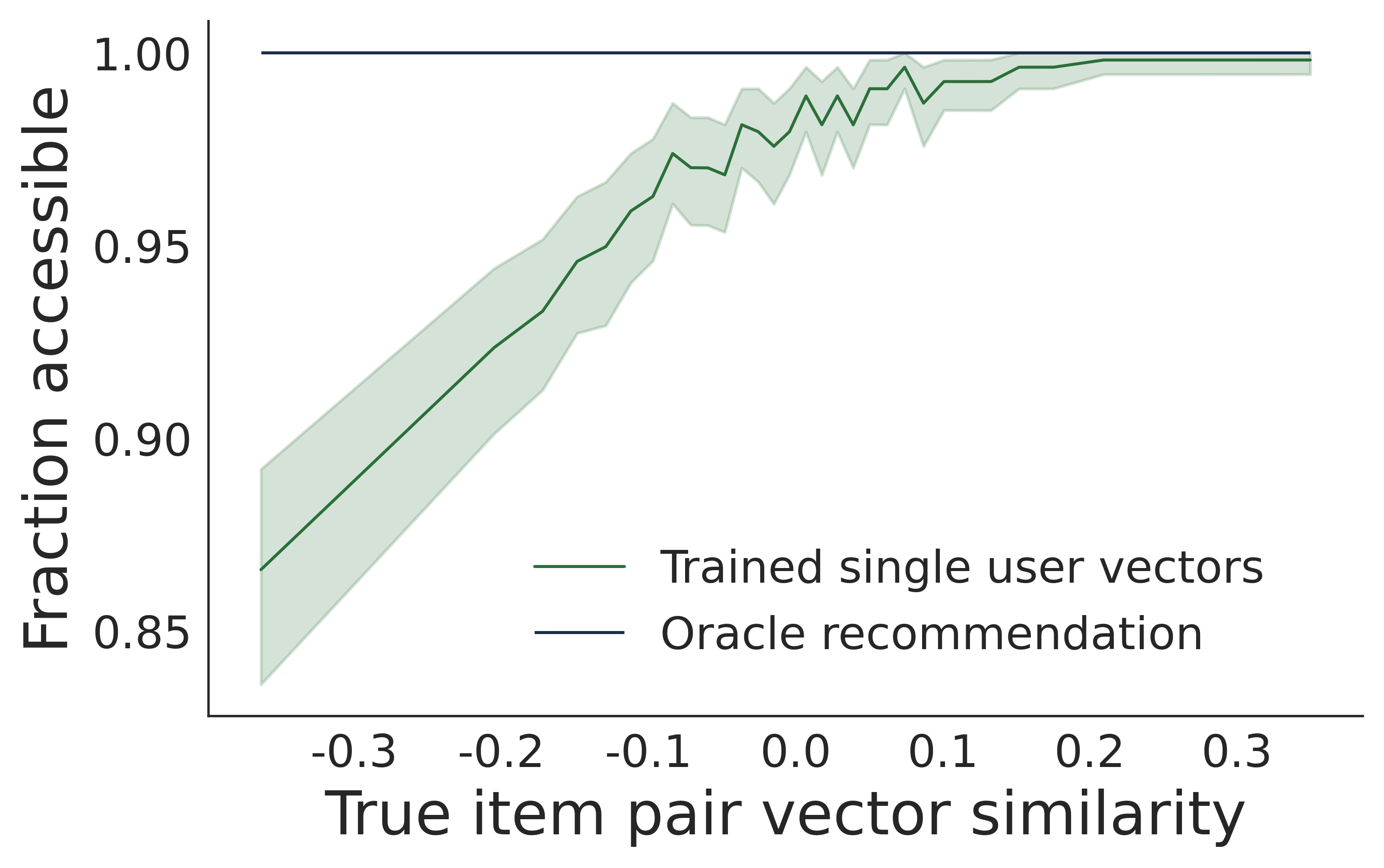}
		\label{fig:trainedimpossibility}}
	\qquad
	\subfloat[Subfigure 3 list of figures text][\rev{Fraction of} item pairs from each topic pair \rev{that} are recommended together for a user vector constructed specifically to maximize the ratings for that pair. ]{
		\includegraphics[height=0.20\linewidth]{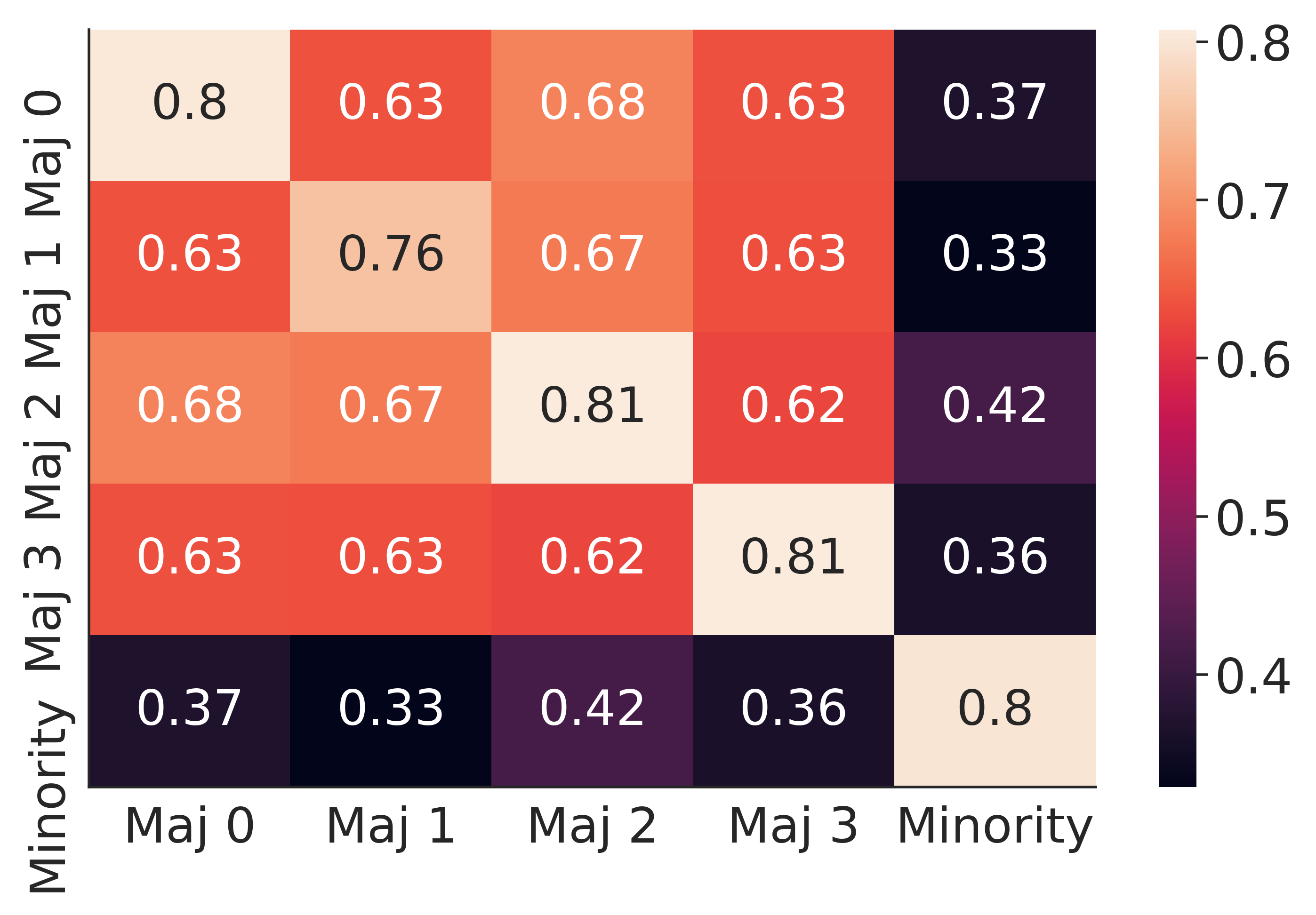}
		\label{fig:syn2minorityim}}
	
	\caption{Empirical accessibility results with two synthetic datasets.}
	
	\Description{The left subfigure is a line graph with two lines, one for the oracle recommendations, and the other one for the trained single user vector recommendations. The y axis shows the average number of recommendations the item pairs with that similarity, ranging from 0 to 10, and the x axis shows the true item pair vector similarity, ranging from -0.4 to 0.4. The recommendation frequencies for both two lines increase with the item pair similarity, where the line for the trained single user vector recommendations increases faster. This shows that the recommendation frequency of item pairs for the trained system is more sensitive to the item similarity. 
	The middle subfigure is a line graph with two lines, one for the oracle recommendations, and the other one for the trained single user vector recommendations.  The y axis shows the fraction of accessible item pairs within all item pairs with a certain item pair similarity, starting from 0.85 to 1. The x axis shows the true item pair vector similarity, starting from -0.4 to 0.4. The fraction of accessible item pairs for the line for the trained single user vector recommendations increases with the item pair similarity, while the oracle line is always at 1. This shows that the accessibility of item pairs for the trained system is highly dependent on the item similarity, where more similar items are much more likely to be joint-accessible. The right subfigure is a heat map, with five rows and five columns. The five groups include four majority groups of users and one minority group of users, corresponding to the dataset where 20 percent of users are  selected as the minority users. Recall that for each majority or minority user, we generated the user feature vectors from a topic model where each user prefers one or two topics. Each cell in the heat map then represents the fraction of item pairs which are preferred by that user as ground truth that are actually recommended to that user group. The majority group of users all have a fraction ranging from 0.6 to 0.81, while the minority users only have fractions ranging from 0.33 to 0.42. This shows that the system in particular fails to capture the minority users' ground truth interests.}
	
	\label{fig:empsyn} \vspace{-7mm}
\end{figure}

First, we observe that the single-vector model does not effectively make item pairs accessible to users. Even restricting the attention to 400 items (and thus \num{79800} unique item pairs), the users combined were only recommended \num{28451}---$35.7\%$---of the item pairs. (Even as all 400 items were recommended individually). In contrast, an oracle multi-vector recommender would show these users \num{42729}---$53.5\%$---unique item pairs (note that if every user was shown a different pair, we would see \num{69878} unique pairs).
Figure~\ref{fig:empiricalappearanceprob} further shows the relationship between how similar two items are, and how \textit{often} they're empirically recommended to users in a top-2 recommendation in the dataset.  On average, using a single-vector recommender, the most similar items are recommended together more than 10 times more often than are the least similar items. In contrast, using the oracle recommender -- which uses multiple vectors per user -- dissimilar items are recommended together almost as often as are similar items. 

The above measures are based on the users present in our synthetic dataset (recall that we sample the user vectors uniformly at random). On the other hand, our theoretical results are whether there exists \textit{any} user vector such that a given pair of items can be recommended together as the top two items. To demonstrate such inaccessibility on our synthetic dataset, for each pair of items we (approximately) find the user vector \textit{best} suited to recommend those two items, using the following heuristic. We wish to find a user vector that maximizes the ratings for the given items while minimizing the ratings of other items; thus, we solve the following least squares problem, for item pair $j,k$: 
\begin{align}\label{eqn:theoraccesslsheuristic}
    u^* &= {\arg\min}_{u} \| \mathbf{V} u - r \|,\,  
    \text{ where}\,\,\,\,\,\,\,\,\,
    r_\ell = \begin{cases} 1 & \ell \in \{j, k\} \\
    0 & \text{otherwise}
    \end{cases} 
\end{align}
and $\mathbf{V}$ is the learned set of item vectors. In other words, we construct a user vector $u^*$ for a hypothetical user that likes items $j,k$ and nothing else. Then, we check whether the given two items are indeed recommended to the user represented by the calculated user vector with top-2 recommendations. A pair is declared inaccessible if indeed they are not the top two items, even for a user constructed especially for these two items. Figure~\ref{fig:trainedimpossibility} shows the results. We find that for a dissimilar pair of item vectors, indeed they are sometimes declared inaccessible by our heuristic; even if a user likes those two items and no other items, that user is not recommended those two items with single-vector matrix factorization.  Recall that, theoretically, there does exist multi-vector representation per user such that every pair would be accessible -- and so an oracle recommender would achieve perfect accessibility according to this metric.

The above metrics are \textit{item} centered: which pairs of items can or cannot be recommended together. However, recall that a core insight of our work is that users with joint interests rare in the population -- even if their individual interests are all popular -- will necessarily receive recommendations that do not match their diverse interests. We now turn to a user-centric measure, in Appendix \Cref{fig:similarityuser}. Consider how diverse a given user's interests are, compared the general population: as measured by the dot product of their two true user item vectors. We find that the more diverse their true interests, the more diverse are their recommendations by the oracle recommender (similarity of recommendations is increasing in similarity of user vectors). On the other hand, with a single user trained vector, users with diverse interests do not receive correspondingly diverse recommendations. Furthermore, recommendations using single-vector model are overall far more similar than are recommendations with the oracle recommender.

Appendix~\ref{app:exp-additional-details} contains the corresponding set of results for $64$-dimensional user vectors. In general, we find that empirical accessibility (with randomly drawn users) results do not change with the dimension (even up to $128$-dimensional user vectors) -- in practice, with a standard training algorithm for single vectors, joint accessibility does not improve with the dimension, even as error metrics such as RMSE do -- in other words, over-parametrization does not help in practice. On the other hand, our heuristic to determine theoretical accessibility as in \Cref{eqn:theoraccesslsheuristic} does predominantly find suitable user vectors -- suggesting that increasing the dimension alongside finding new methods to train user vectors may also increase accessibility in practice. 

\paragraph{Second synthetic setting.} 

\rev{We turn our attention to the particular effects on minority users, using our second synthetic dataset. Results similar to those presented in \Cref{fig:empml} are in the this data setting as well (to even greater magnitudes, due to the effect of topics), and are in \Cref{app:exp-additional-details}. However, we see further effects on the minority users. \Cref{fig:syn2minorityim} shows, for each pair of topics, the fraction of item pairs with those topics that are accessible via the heuristic used for \Cref{fig:trainedimpossibility}. While most of the majority-topic items are jointly accessible, even across topics, only about 35\% of Minority-Majority item pairs are recommended -- even to a user who likes that pair of items and nothing else. Such inaccessibility further leads to stereotyping of minority users in the dataset. Recall that individual minority users, by design, equally like items from the minority topic and one of the majority topics. However, using single-item representation, they are almost exclusively recommended items only from the minority topic: over 99\% of items recommended via top-2 recommendation to minority users are from the minority topic -- whereas using the ground truth, only about 45\% would be. Finally, such users are \textit{never} shown items from multiple topics with top-2 recommendation; whereas using the ground truth, 30\% of pairs recommended would come from different topics. These results stem from the effects described above: learned item vectors for the minority topic are far from those learned for the other topics, and single-vector user representation cannot be simultaneously close to items from both minority and majority topics.}  

\paragraph{MovieLens dataset.}

Appendix \Cref{fig:empml} contains the results for ML10M, where we do not have an oracle recommender system against which to compare anymore. Results are qualitatively similar to those using our first synthetic dataset, or even more extreme. Only \num{6433}---$8.1\%$---of pairs are recommended together to the training set users, and the most similar items are recommended together almost $100$ times more often than the least similar items. (Even as $89\%$ of the items can be individually recommended). Restricting among items that are individually recommended, for most pairs of items we can find no user vector that recommends that pair together. Together, these empirical results demonstrates that joint accessibility -- especially of items deemed dissimilar by the user population -- is a challenge in practice.







\section{Conclusion} \label{sec:conc}

We have studied a stereotyping problem in matrix factorization-based collaborative filtering algorithms with regard to  users' accessibility to a diverse set of items -- standard recommenders cannot serve users with joint interests that are anti-correlated in the general population. We formalize this challenge by introducing a new notion of \textit{joint accessibility}, which describes whether a set of items can \textit{jointly} be accessed to a user in a recommendation and its geometric interpretations. Such a notion can be violated when the user is only represented in a single-vector representation model. We further propose an alternative modelling approach, which is designed to capture the diverse multiple interests of each user using a \textit{multi}-vector representation. Extensive experimental results on real and simulated datasets demonstrate the stereotyping problem with standard single-vector matrix factorization models.

As recommender systems are increasingly employed in numerous online platforms and high stakes environments, it is necessary to scrutinize to ensure that these systems will not perpetuate, exacerbate, or create new inequity problems. Aiming to make recommender algorithms themselves intrinsically fairer, more inclusive, and more equitable plays an important role in achieving that goal. In this work, we inspect one aspect of such impact, which is on the stereotyping problem of diverse user interests, and provide new analysis on how the joint accessibility of items can be limited in a standard matrix factorization-based recommender system. More generally, our analysis emphasizes the importance of model and algorithm choice to the system's downstream societal impact.





\bibliographystyle{ACM-Reference-Format}
\bibliography{ref}

\clearpage
\appendix

\section{Proofs}

\subsection{Proof of Theorem \ref{prop:single-vec-vertex-condition}} \label{app:proof:thm:single-vec-vertex-condition}

\begin{proof}
The proof is a generalized version of \citet[Proposition~1]{dean2020recommendations}, which is specialized to Top-1 recommendation. First, notice that, a set $S$ of $K$ items is recommended as the set of Top-$K$ items if and only if: 
\begin{align*}
    \sum_{i\in S} \vv_{i}^\top u > \sum_{i\in S'} \vv_{i}^\top u,
\end{align*}
for all $S' \subseteq [n], S' \neq S, |S'| = K$. The above is equivalent to:
\begin{align}\label{eq:single-vec-top-k}
    Q_s u > 0,
\end{align}
where $Q_s = \begin{pmatrix} (\sum_{i\in S} \vv_{i} -  \sum_{i\in S'_1} \vv_{i})^\top\\ \cdots\\ \cdots \\ \left(\sum_{i\in S} \vv_{i} -  \sum_{i\in S'_{{n\choose k}-1}} \vv_{i}\right)^\top\end{pmatrix} \in \br^{\left({n\choose k}-1\right) \times d}$.
Therefore, the set $S$ is accessible if and only if there exists $u \in \br^d$, s.t, eq.~\eqref{eq:single-vec-top-k} is satisfied. This is a linear constraint on $u$, and is equivalent to the feasibility of the following linear program:
\begin{align*}
    \min  \quad & \zero^\top u \\
    \st \quad & Q_su \geq \epsilon
\end{align*}
where $\epsilon>0$ is an arbitrarily small positive number. To analyze the feasibility of this linear program, we consider its dual:
\begin{align*}
    \max  \quad & \epsilon^\top \lambda \\
    \st \quad & Q_s^\top \lambda =0\\
    &\lambda \geq 0
\end{align*}
Since the dual is always feasible with a solution $\lambda = \zero$, by Clark's theorem, the primal is feasibly if and only if the dual is bounded. Further notice that for any $\lambda$ that is a feasible solution, $a\lambda, a\geq 0$ is also a feasible solution. Setting $a$ to $+\infty$ will make the dual unbounded whenever if $\lambda$ has nonzero entries. This means that a sufficient condition for the unboundedness of the dual is that there exists some feasible $\lambda \neq \zero$, with $Q_s^\top \lambda =0$, such that:
\begin{align*}
    \sum_j \lambda_j \left(\sum_{i\in S} \vv_{i} -  \sum_{i\in S'_j} \vv_{i}\right) = 0, \lambda > 0.
\end{align*}

By rearranging terms, this is equivalent to $\sum_{i\in S} \vv_{i}$ being a convex combination of $ \sum_{i\in S'_j} \vv_{i}$ of all $j$. Therefore, $S$ is accessible for a user $u$ if and only if $\sum_{i \in S} \vv_i$ is a vertex of the convex hull: $\conv\{\sum_{i \in S} \vv_i, \forall S \subseteq [n], |S| = K\}$.
\end{proof}

\subsection{Proof of Theorem~\ref{thm:voronoi}}\label{app:proof:thm:voronoi}

\begin{proof}
First, assume that items $\vv_1$ and $\vv_2$ are the top-2 recommendations for some user $\vu$. Then by definition, for all $j \in [n]$, $j \neq 1, 2$, we have $\vu \cdot \vv_j \leq \vu \cdot \vv_1$, and $\vu \cdot \vv_j \leq \vu \cdot \vv_2$. Without loss of generality, assume that $\vv_1$ is the Top-1 recommendation for $\vu$. Then by the construction of Voronoi diagram, this means that $\vu$ is in the Voronoi cell of $\vv_1$. 

Suppose that if $\vv_2$ is not in an neighboring Voronoi cell of $\vv_1$'s. Then, there must exist another item vector $\vv_3$, such that $\vv_3$ has a neighboring Voronoi cell to $\vv_1$, and the closest distance (the shortest arc) between $\vv_1$ and $\vv_2$ crosses the cell of $\vv_3$. Denote the shortest arc from $\vv_1$ to $\vv_2$ on the sphere as $\ell$.

If $\vv_3$ is on $\ell$, we immediately get $d(\vu, \vv_3) < d(\vu, \vv_2)$. If $\vv_3$ is not on $\ell$, consider any point $\vv_4$ that is in the Voronoi cell of $\vv_3$ and is on $\ell$. Then, by triangle inequality, 
\begin{align*}
    d(\vu, \vv_3) &< d(\vu, \vv_4) + d(\vv_4, \vv_3)\\
    & < d(\vu, \vv_4) + d(\vv_4, \vv_2)\\
    &= d(\vu, \vv_2).
\end{align*}
Therefore, $\vu$ is closer to $\vv_3$ than $\vv_2$. This contradicts the fact that $\vv_2$ is the second closest item vector to $\vu$.

On the other hand, consider two item vectors $\vv_1$ and $\vv_2$ that are Voronoi neighbors. Then, any user vector $\vu$ that is on the intersection of the two Voronoi cells has a top-2 recommendation of $\{\vv_1, \vv_2\}$, which completes the proof.
\end{proof}

\subsection{Proof of Theorem~\ref{thm:single-vec-impossibility}}\label{app:proof:thm:single-vec-impossibility}

\begin{proof}
We prove the theorem by providing an example that is illustrated in Figure~\ref{fig:convexhull-voronoi} \textit{(left)}. Specifically, consider $d=2$, and four items with feature vectors $(2,4), (-2,2), (-3, -1), (3,-3)$. It is immediate to see that for top-1 recommendations, four item vectors are all vertices of the convex hull $\conv\{\sum_{i \in S} \vv_i, \forall S \subseteq [n], |S| = 1\}$. Therefore, by Theorem~\ref{app:proof:thm:single-vec-vertex-condition}, each item is individually accessible.

Now we consider a top-2 recommendation. The convex hull $\conv\{\sum_{i \in S} \vv_i, \forall S \subseteq [n], |S| = 2\}$ is a convex hull of six possible item pairs, which correspond to the vector sums: \[(0,6), (-1,3), (5,1), (-5, 1), (1,-1), (0,-4).\] Therefore, the item pair of item 1 and item 3 (correspond to vector sum $(-1,3)$),  and the item pair of item 2 and item 4 (correspond to vector sum $((1,-1))$) are not jointly accessible. Therefore, with the single vector recommendation, individual accessibility of the items is not sufficient for joint accessibility. This completes the proof.
\end{proof}

\subsection{Proof of Theorem \ref{thm:multi-vec-access}} \label{app:proof:thm:multi-vec-access}

\begin{proof}
First, given that each item is individually accessible, by Theorem~\ref{app:proof:thm:single-vec-vertex-condition} with $K=1$, we have that for any item $j \in [n]$, there exists a user vector $\vu_j \in \R^d$, such that
\[
\vu_j  \cdot \vv_j \geq \vu_j  \cdot \vv_{j'},
\]
for any $j' \in [n], j' \neq j$. For convenience, let us denote $\vu_j$ as the representative user vector for item $j$.

Moreover, consider an arbitrary set of $K$ items, with item feature vectors $\vv_1, \cdots, \vv_K$ and their representative user vectors $\vu_1, \cdots, \vu_K$. Then there exist constants $c_1, \cdots, c_k$ such that
\[
c_1\vu_1\cdot \vv_1 = \ldots = c_K\vu_K\cdot \vv_K.
\]
Now we show that the set of user vectors $\{c_i\vu_i\}, i = 1\cdots K$ form a multi-vector user representation such that items $\cS = \{\vv_1, \cdots, \vv_K\}$ are jointly accessible. 

Consider any other set of $K$ items with item feature vectors  $\cS' = \{\vv'_1, \cdots, \vv'_K\}$. Denote the predicted ratings for set $\cS$ and $\cS'$ as $r(\cS)$ and $r(\cS')$. The scoring rule as in Eq~\eqref{eq:score-multi} yields:
\begin{align*}
    r(\cS) &= \sum_{i=1}^K \max_{j \in [m]} c_j\vu_j\cdot \vv_i= \sum_{i=1}^K  c_i\vu_i\cdot \vv_i \geq \sum_{i=1}^K \max_{j \in [m]} c_j\vu_j\cdot \vv'_i = r(\cS'),
\end{align*}
which completes the proof.

\end{proof}

\section{Additional Experimental Details and Results}\label{app:exp-additional-details}

\subsection{Synthetic data generation and training details} 

\paragraph{Data generation details.}

In the first synthetic setting, we sample \num{10677} items and  \num{69878} users, matching the size with the ML10M dataset. As the ground truth, we represent each item as a \num{64} dimensional vector uniformly sampled from the unit sphere. To illustrate exactly how single-vector representation fails when users might have different (potentially ``opposing'' interests), we construct for each user a ground truth representation as two \num{64} dimensional vectors uniformly sampled from the unit sphere. That is,
\[
  \vu_{i, k} = \frac{Z_{i, k}}{\|Z_{i, k}\|_2} \quad
  \vv_j = \frac{Z_j}{\|Z_j\|_2},
\]
where $\vu_{i, k}$ is the $k$-th (for $k\in \{1, 2\}$) vector representing user $i$ and $\vv_j$ is the vector representing item $j$, and $Z_{i, k}$ and $Z_j$ are vectors drawn from $\mathcal{N}(0, I_d)$. Then, the corresponding true rating between a user-item pair is 

\[r_{i, j} = \max_{k \in \{1, 2\}}\langle \vu_{i, k}, \vv_j \rangle.\]

For training data, we uniformly sample 10 million user-item pairs without replacement and obtain their corresponding rating, with an additive Gaussian noise of mean zero and standard deviation \num{0.01}.

For the second synthetic setting, we first draw five 64 dimensional topic vectors on the unit ball. To force the Minority topic to be anti-correlated with the others, we force the first 20 dimensions of the Majority topic to be positive and of the Minority topic to be negative. Then, for each item, we first sample a topic uniformly at random and then add independent and identically distributed gaussian noise to each dimension, with standard deviation $0.05$. For each user, we first draw 2 topics (with replacement) uniformly at random (except with limits for majority and minority users as described in the main text) and then add independent and identically distributed gaussian noise to each dimension of each vector, also with standard deviation $0.05$.


\paragraph{Further training details.}

For all three datasets, we train matrix factorization models using the standard alternating least squares (ALS) algorithm \citep{hu2008collaborative}. In each setting, we model each user and item as a single vector of dimension $32$ or $64$. We tune the regularization parameter using $10$-fold cross validation and grid search. In the synthetic setting, we further constrain vectors to have norm $1$, to match with the ground truth factors.
We project all item vectors to the unit ball, and ignore item bias terms.\footnote{Doing so allows us to focus on \textit{joint} accessibility -- recall that shown by \citet{dean2020recommendations}, individual items may not be accessible if, for example, they have small norm or negative bias terms.}

\subsection{Additional experimental results}

\begin{figure}[!ht]
	\centering
			\subfloat[Subfigure 1 list of figures text][How similar the top two items recommended to a user are to each other, as a function of how similar the two users true vectors are.]{
		\includegraphics[width=0.445\linewidth]{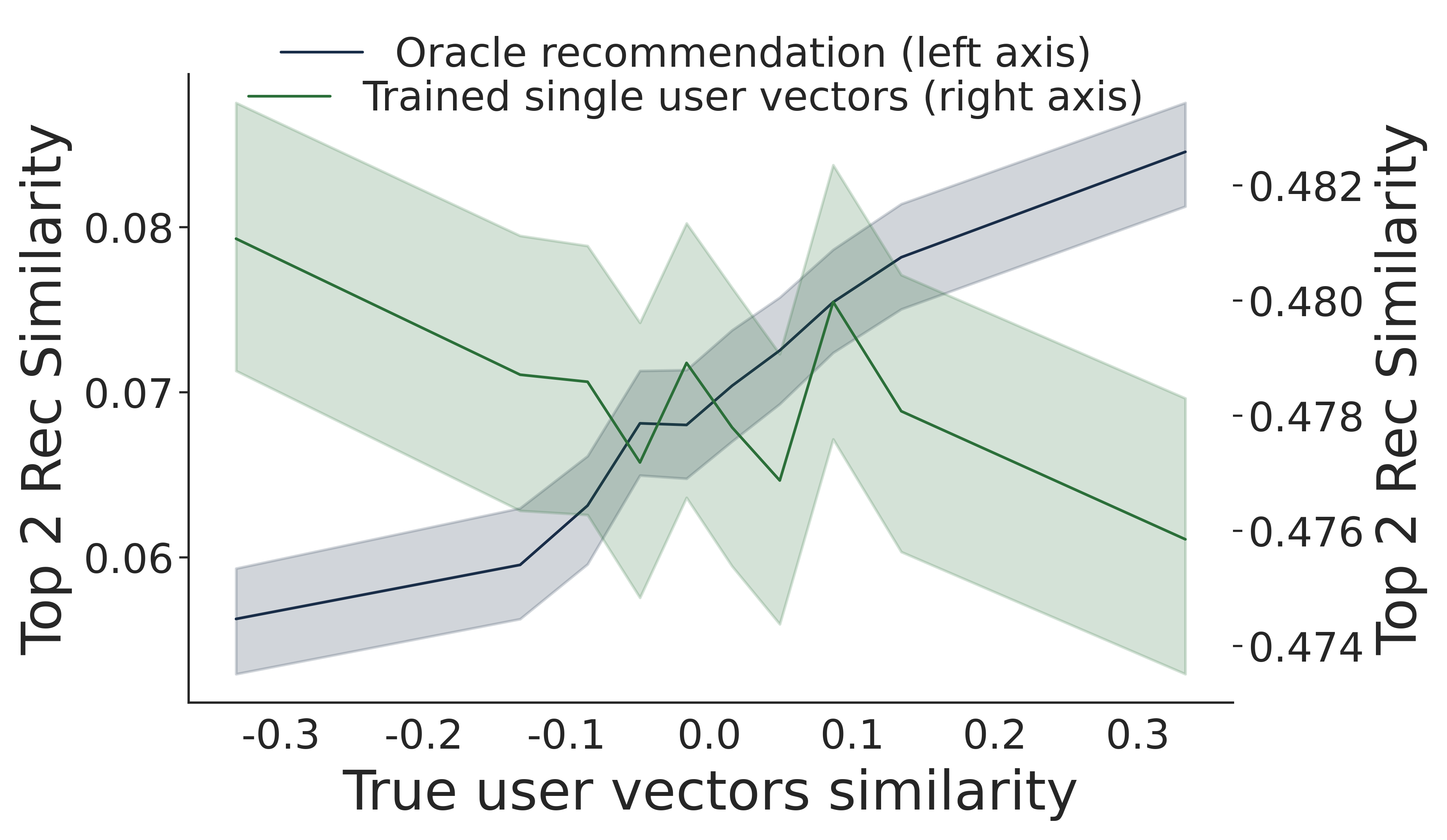}
		\label{fig:similarityuser}	}
			\qquad
	\subfloat[Subfigure 1 list of figures text][Same plot as~\Cref{fig:empiricalappearanceprob}, with 64-dimensional vectors]{
		\includegraphics[width=0.445\linewidth]{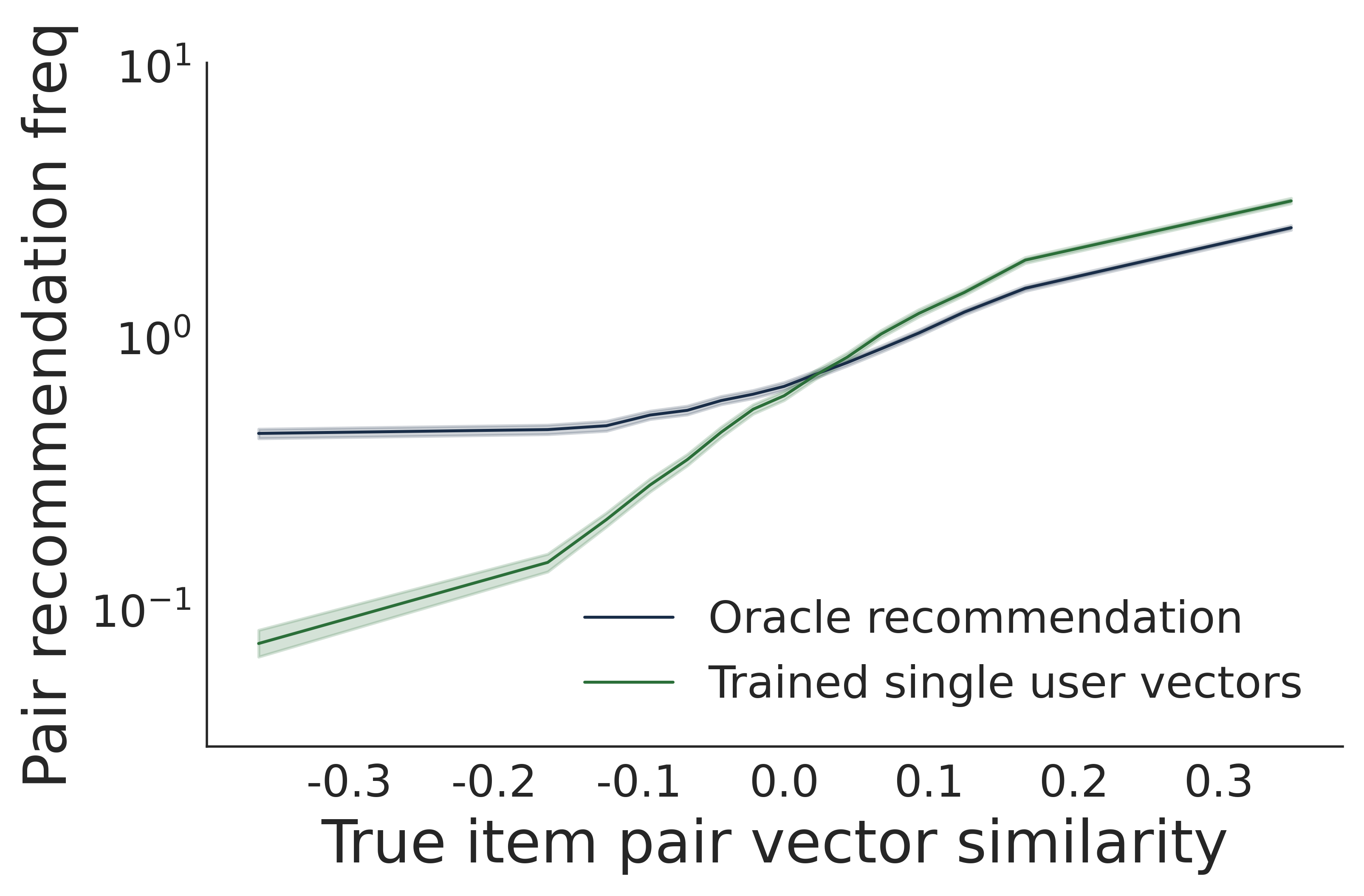}
		\label{fig:syn64empir}
		
		\Description{The right subfigure is a line graph with two lines, one for the oracle recommendations, and the other one for the trained single user vector recommendations using 64 dimensional features. The y axis shows the average number of recommendations the item pairs with that similarity, ranging from 0 to 10, and the x axis shows the true item pair vector similarity, ranging from -0.4 to 0.4. The recommendation frequencies for both two lines increase with the item pair similarity, where the line for the trained single user vector recommendations increases faster. This shows that the recommendation frequency of item pairs for the trained system is more sensitive to the item similarity. The left subfigure is a line graph with two lines, one for the oracle recommendations, and the other one for the trained single user vector recommendations. The y axis shows the similarity score between the top two items recommended to a user, starting from 0.05 to 0.10. The x axis  shows the true item pair vector similarity, ranging from -0.4 to 0.4.  For the oracle recommendations, the line is increasing; while for the trained recommendations, the line is increasing. This shows that even for the oracle recommendations the top two items are not necessarily need to be more similar, the trained system has such effect.}}

	\caption{Supplemental figures for the first synthetic experiment.}
	\label{fig:empsyn64}
\end{figure}

In this section we include further empirical results for the synthetic and ML10M datasets. Figure \ref{fig:empsyn64} shows the accessibility of items in the synthetic setting with a $64$ dimensional matrix factorization model. Figure \ref{fig:empml64} shows the accessibility of items in the ML10M case with a $64$ dimensional matrix factorization model. 

\rev{\Cref{fig:empmlrandomitems} shows the same results except with a random subset of \num{400} items, as opposed to the most common items. When items have orders of magnitude different ratings, however, the norm of the learned embeddings may differ dramatically, and so individual item reachability (cf. \citet{dean2020recommendations}) is now a concern. To highlight the role of joint accessibility, we thus normalize the item vectors, as in our synthetic setting. Finally, \Cref{fig:empmltopitemsnormalized} shows the same findings with the most frequent items (as in the main text), but now with normalized vectors. }

\begin{figure}[!ht]
 \vspace{-3mm}
	\centering
	\subfloat{
		\includegraphics[width=0.36\linewidth]{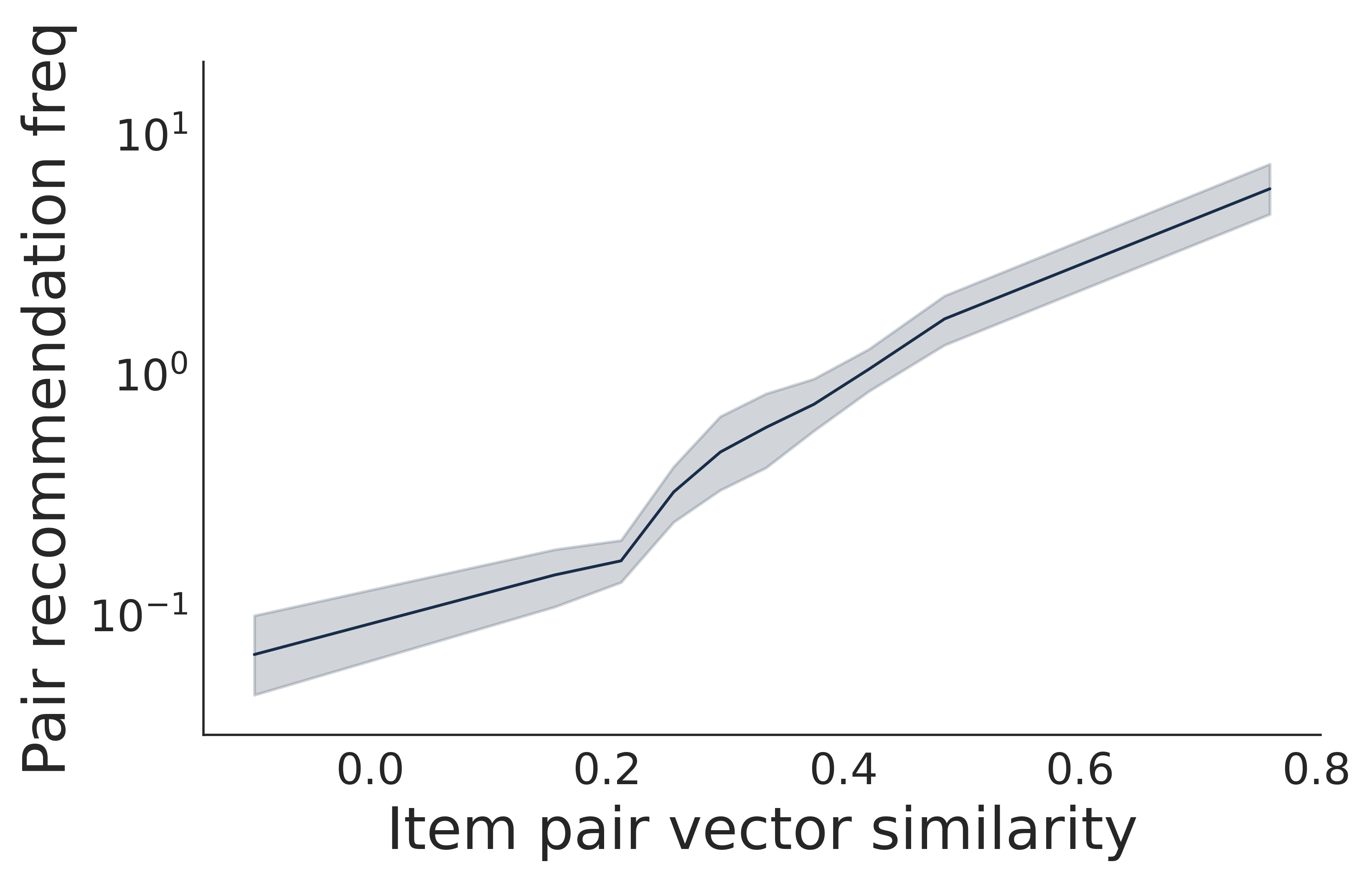}
		\label{fig:empiricalappearanceml}}
	\qquad
	\subfloat{
		\includegraphics[width=0.36\linewidth]{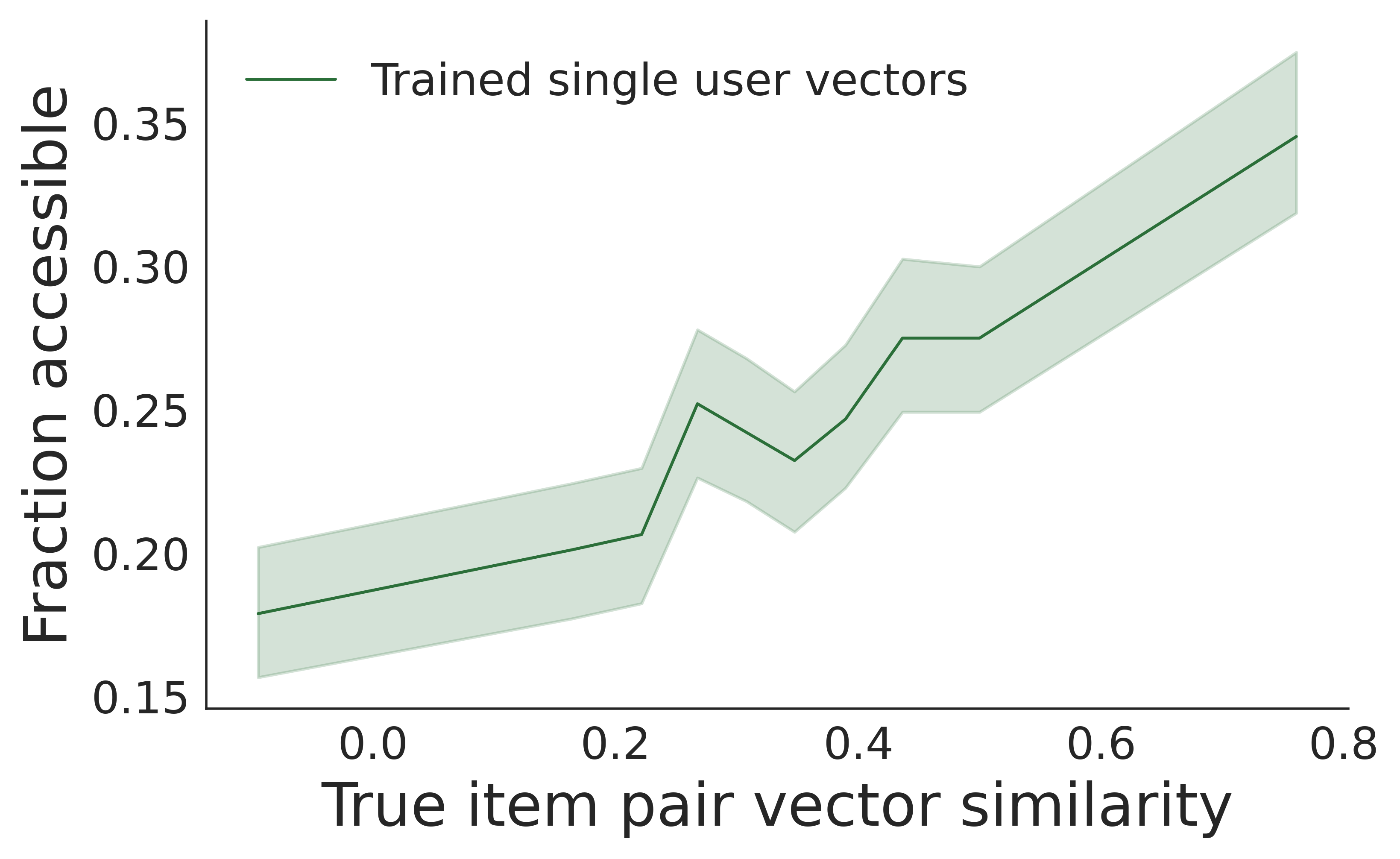}
		\label{fig:trainedimpossibilityml}}
	
	\caption{Empirical accessibility results for ML10M. Each subplot shows the same result as the corresponding subplot in \Cref{fig:empsyn}. Now, vector similarity is defined as cosine similarity, to incorporate that vectors may have differing norms.}
	\label{fig:empml} 
	
	\Description{The left subfigure and right subfigure each is a line graph with one single line: one for the oracle recommendations, and the other one for the trained single user vector recommendations.  In the left plot, the y axis shows the average number of recommendations the item pairs with that similarity, ranging from 0 to 10, and the x axis shows the true item pair vector similarity, ranging from 0 to 0.8. The recommendation frequency increases with the item pair similarity. In the left plot,  the y axis shows the fraction of accessible item pairs within all item pairs with a certain item pair similarity, starting from 0.15 to 0.35, and the x axis shows the true item pair vector similarity, ranging from 0 to 0.8. The accessible fraction increases with the item pair similarity. This shows that more similar items tend to be recommended together, thus more jointly accessible.}
\end{figure}

\begin{figure}[!ht]
	\centering
	\subfloat{
		\includegraphics[width=0.46\linewidth]{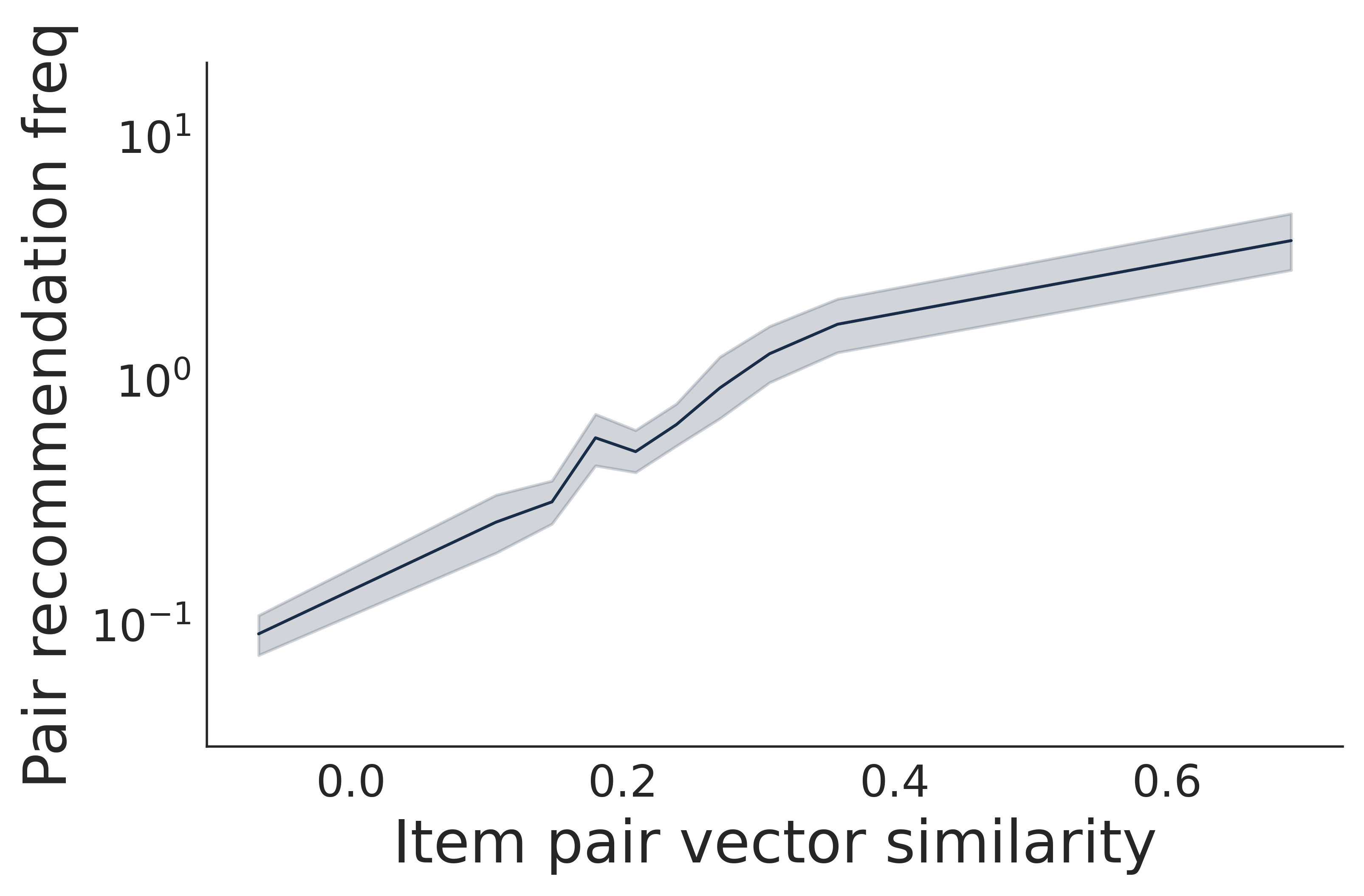}
		\label{fig:empiricalappearanceml64}}
	\qquad
	\subfloat{
		\includegraphics[width=0.46\linewidth]{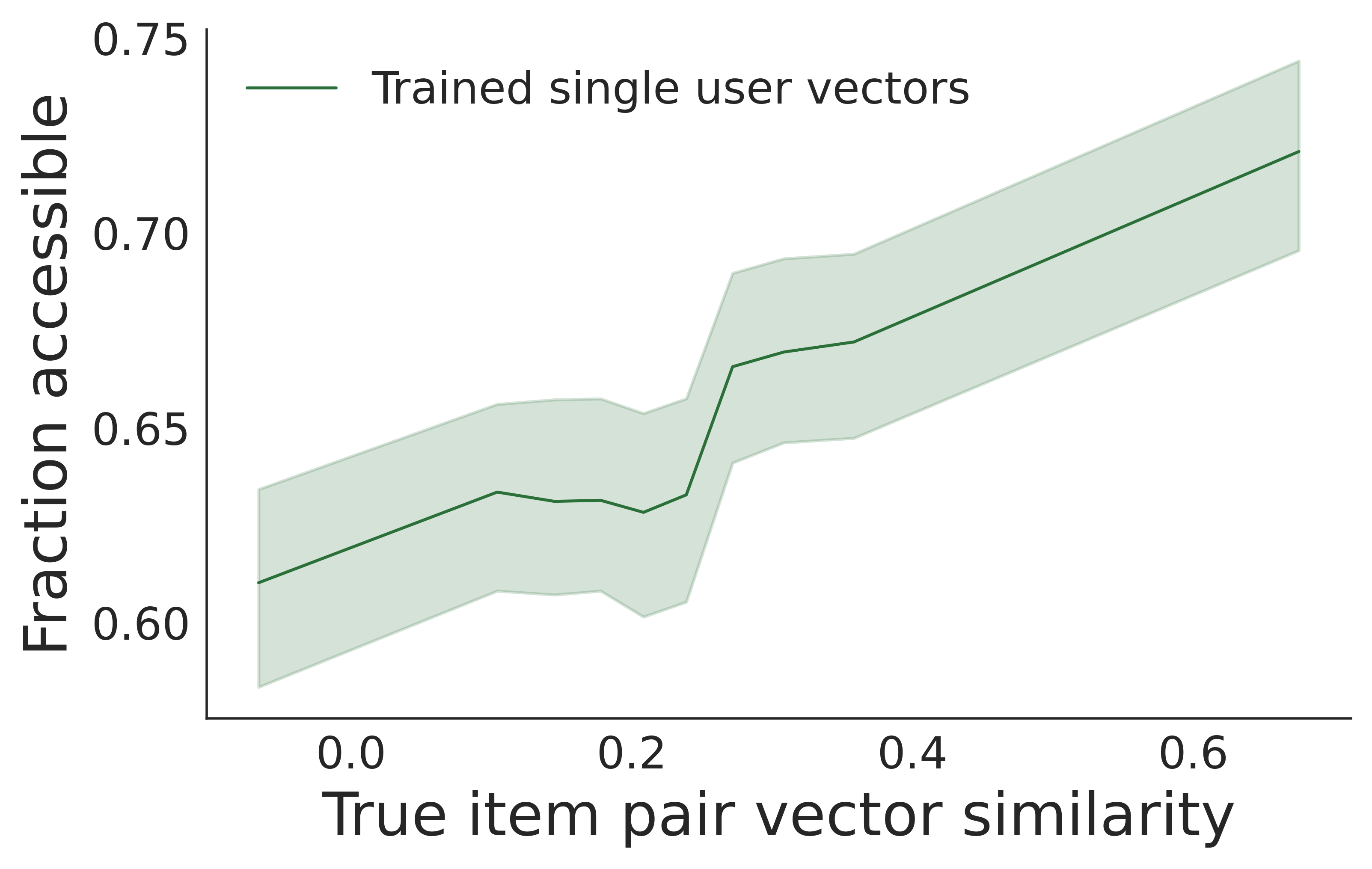}
		\label{fig:trainedimpossibilityml64}}
	
	\caption{Same plots as in \Cref{fig:empml}, except with 64-dimensional vectors.}
	\label{fig:empml64}
	
	\Description{The left subfigure and right subfigure each is a line graph with one single line: one for the oracle recommendations, and the other one for the trained single user vector recommendations with 64 feature dimensions.  In the left plot, the y axis shows the average number of recommendations the item pairs with that similarity, ranging from 0 to 10, and the x axis shows the true item pair vector similarity, ranging from 0 to 0.8. The recommendation frequency increases with the item pair similarity. In the left plot,  the y axis shows the fraction of accessible item pairs within all item pairs with a certain item pair similarity, starting from 0.6 to 0.75, and the x axis shows the true item pair vector similarity, ranging from 0 to 0.8. The accessible fraction increases with the item pair similarity. This shows that more similar items tend to be recommended together, thus more jointly accessible.}
\end{figure}

\begin{figure}[!ht]
	\centering
	\subfloat{
		\includegraphics[width=0.46\linewidth]{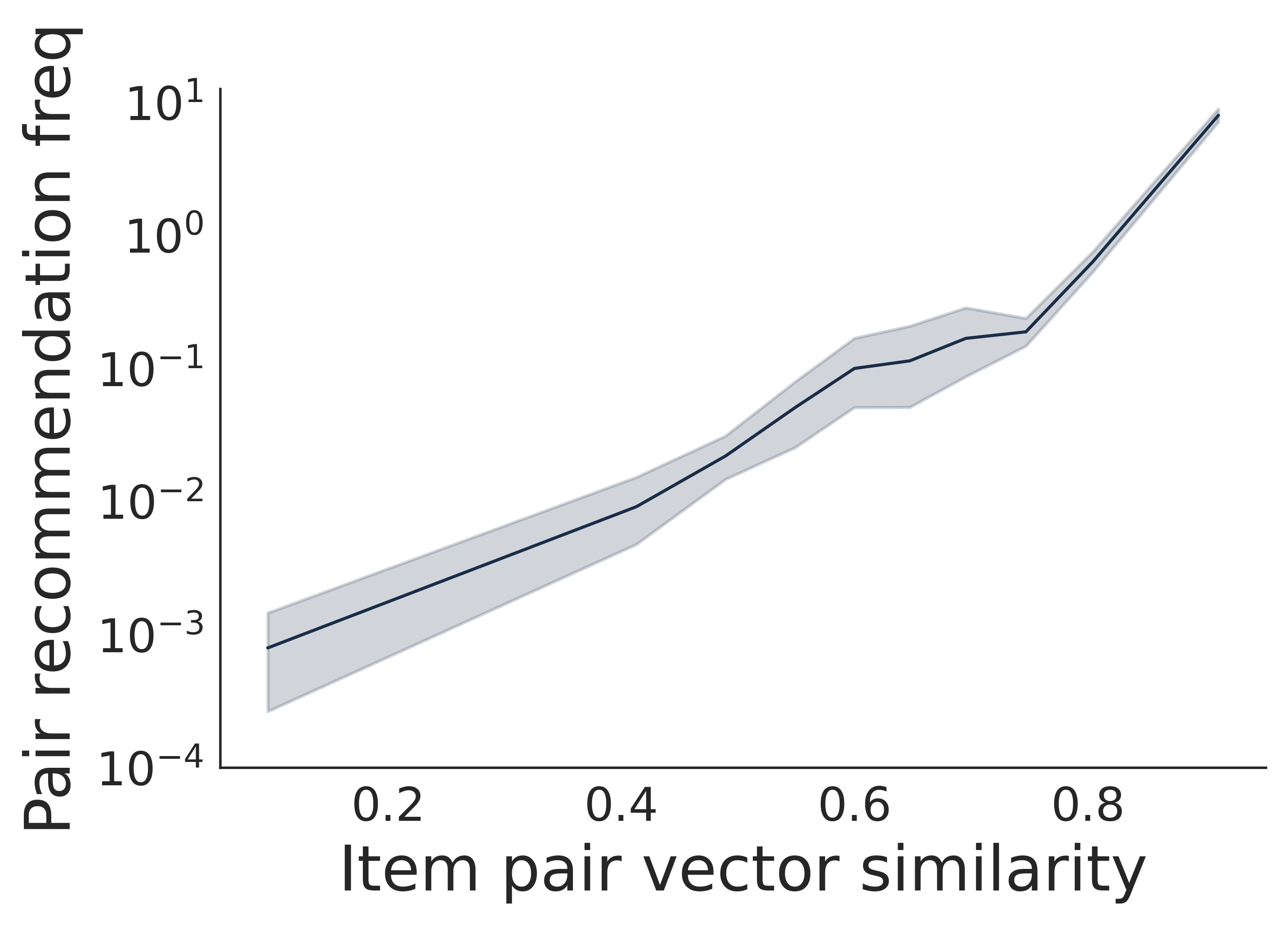}
		\label{fig:empiricalappearancemltop}}
	\qquad
	\subfloat{
		\includegraphics[width=0.46\linewidth]{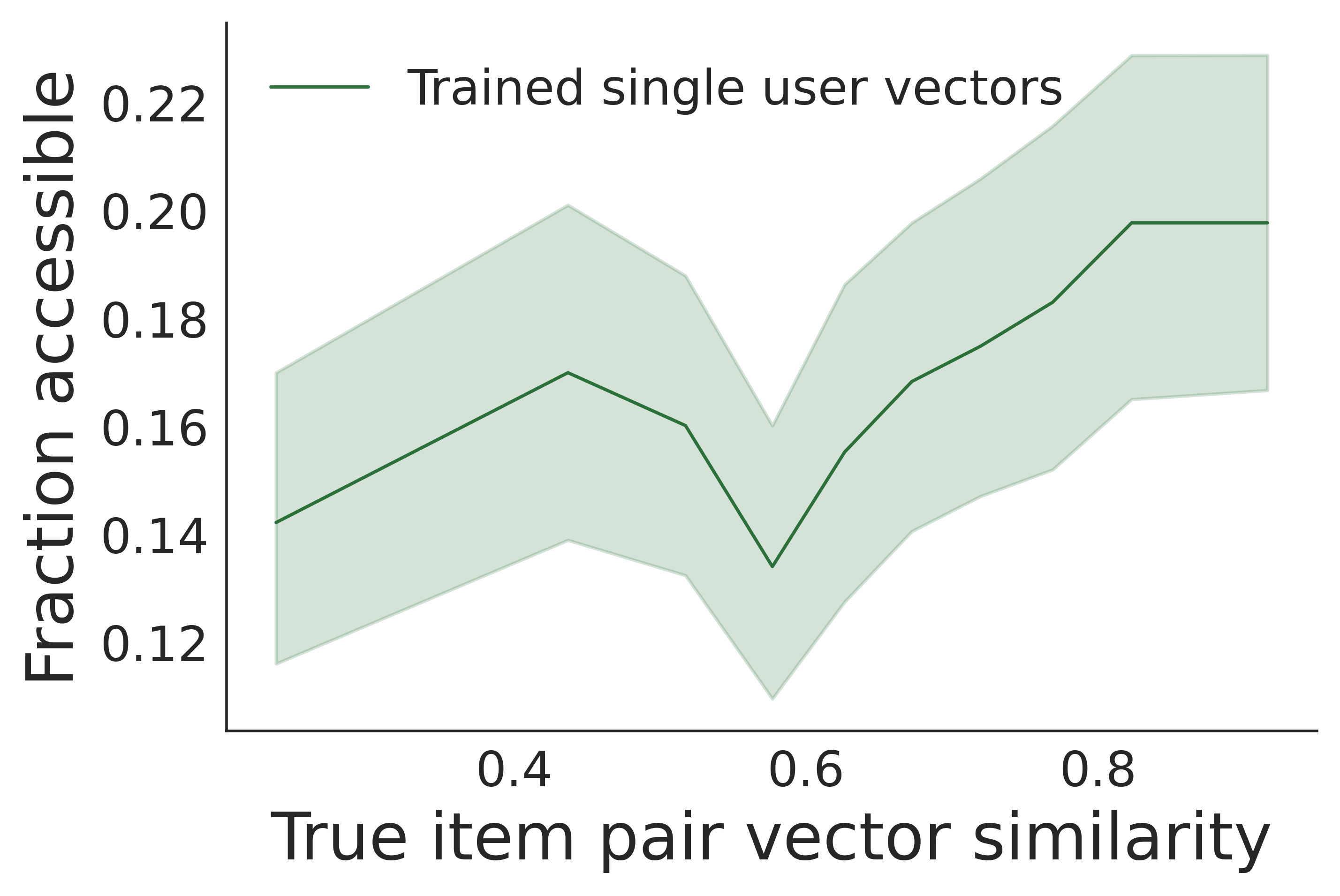}
		\label{fig:trainedimpossibilitymltop}}
	
	\caption{Same plots as in \Cref{fig:empml}, except with randomly chosen items from MovieLens instead of the most frequently rated items (with normalized vectors).}
	\label{fig:empmlrandomitems}
	
	\Description{The left subfigure and right subfigure each is a line graph with one single line: one for the oracle recommendations, and the other one for the trained single user vector recommendations with randomly chosen items from MovieLens 10M dataset.  In the left plot, the y axis shows the average number of recommendations the item pairs with that similarity, ranging from 0 to 10, and the x axis shows the true item pair vector similarity, ranging from 0 to 0.8. The recommendation frequency increases with the item pair similarity. In the left plot,  the y axis shows the fraction of accessible item pairs within all item pairs with a certain item pair similarity, starting from 0.12 to 0.22, and the x axis shows the true item pair vector similarity, ranging from 0.2 to 1. The accessible fraction increases with the item pair similarity. This shows that more similar items tend to be recommended together, thus more jointly accessible in this setting.}
\end{figure}

\begin{figure}[!ht]
	\centering
	\subfloat{
		\includegraphics[width=0.46\linewidth]{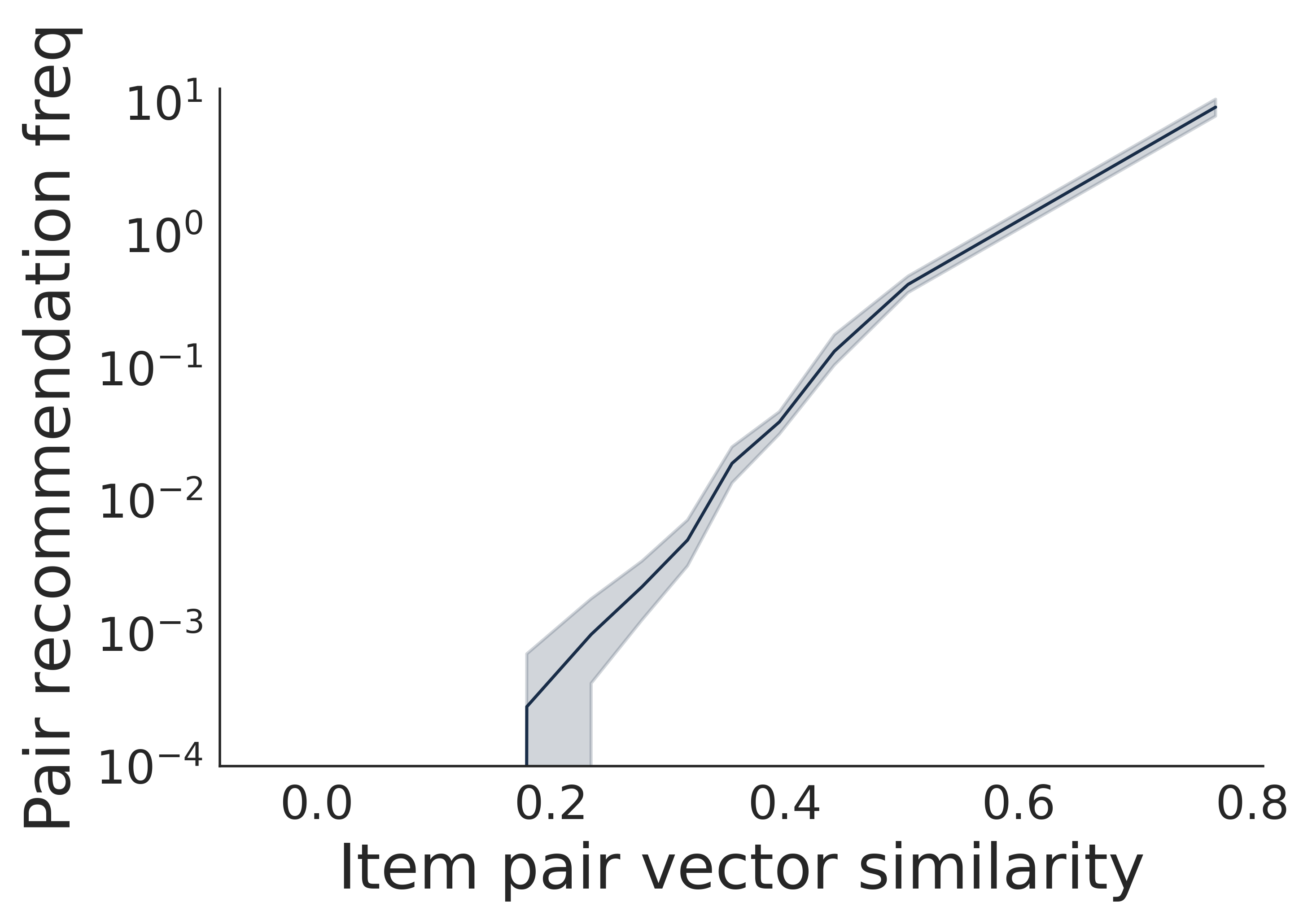}
		\label{fig:empiricalappearancemlnormalized}}
	\qquad
	\subfloat{
		\includegraphics[width=0.46\linewidth]{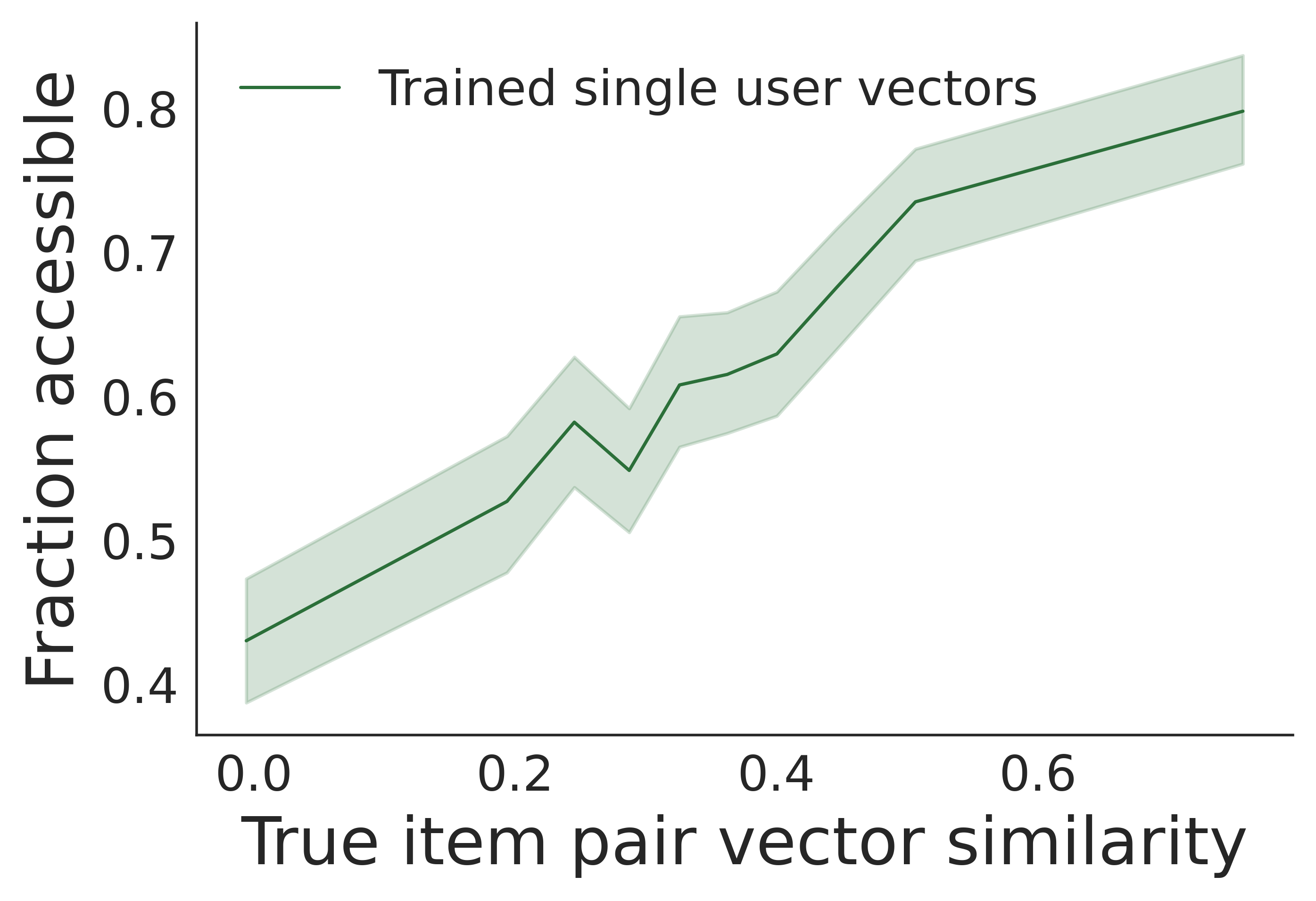}
		\label{fig:trainedimpossibilitymlnormalized}}
	
	\caption{Empirical accessibility results for ML10M. (same plots as in \Cref{fig:empml}, except with normalized item vectors.}
	\label{fig:empmltopitemsnormalized}
	
	\Description{The left subfigure and right subfigure each is a line graph with one single line: one for the oracle recommendations, and the other one for the trained single user vector recommendations with MovieLens 10M dataset and normalized item vectors.  In the left plot, the y axis shows the average number of recommendations the item pairs with that similarity, ranging from 0 to 10, and the x axis shows the true item pair vector similarity, ranging from 0.2 to 0.8. The recommendation frequency increases with the item pair similarity. In the left plot,  the y axis shows the fraction of accessible item pairs within all item pairs with a certain item pair similarity, starting from 0.4 to 0.8, and the x axis shows the true item pair vector similarity, ranging from 0 to 0.8. The accessible fraction increases with the item pair similarity. This confirms that more similar items tend to be recommended together, thus more jointly accessible even with normalized item vectors.}
	
	\end{figure}

The results in all cases are qualitatively similar to Figures \ref{fig:empsyn} and \ref{fig:empml}, which indicates that the stereotyping problem is not merely caused by an underspecified model and that the results are robust.

\begin{figure}[!ht]
 \vspace{-3mm}
	\centering
	\subfloat{
		\includegraphics[width=0.45\linewidth]{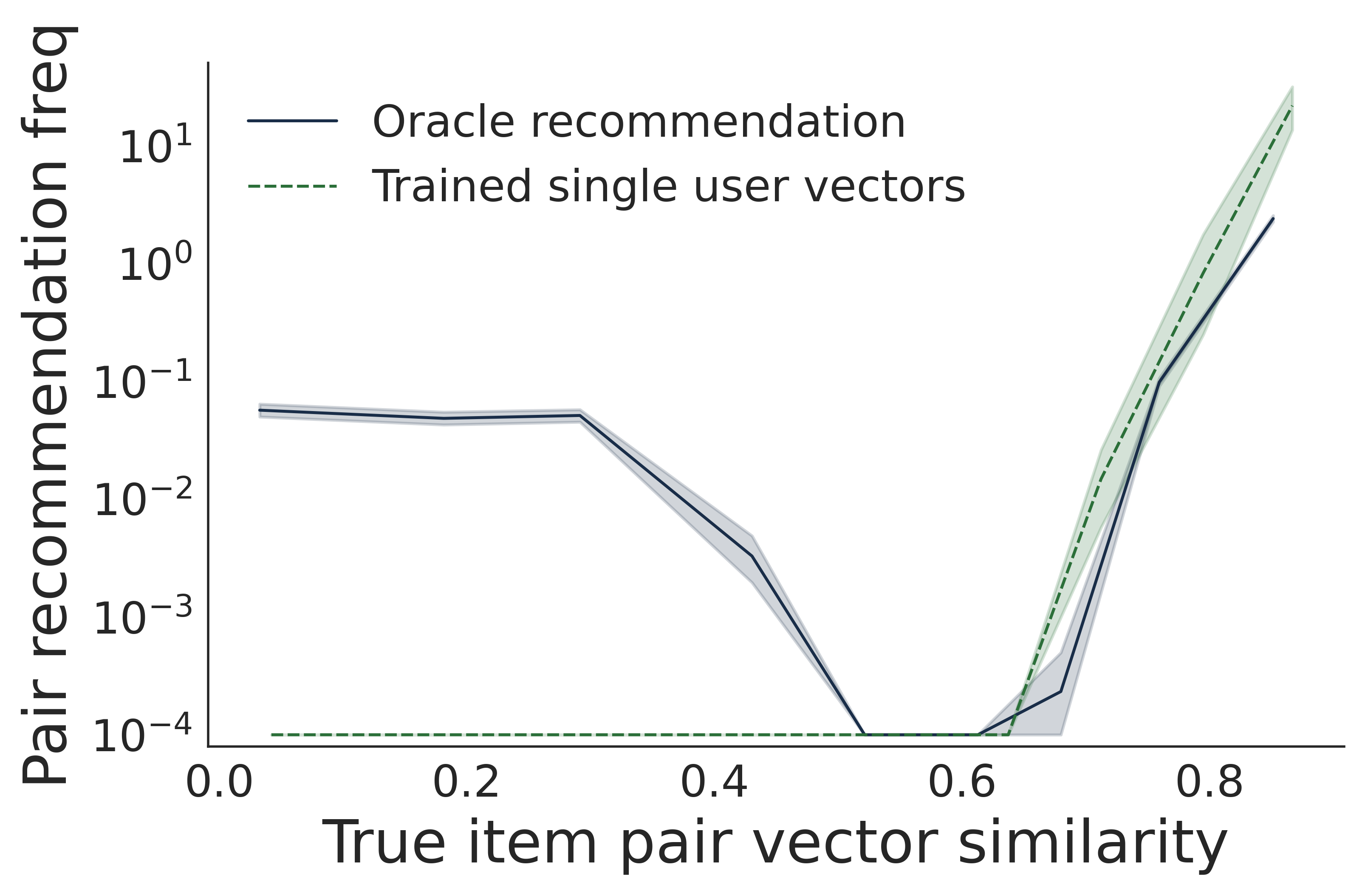}
		\label{fig:empiricalappearancesyn2}}
	\qquad
	\subfloat{
		\includegraphics[width=0.45\linewidth]{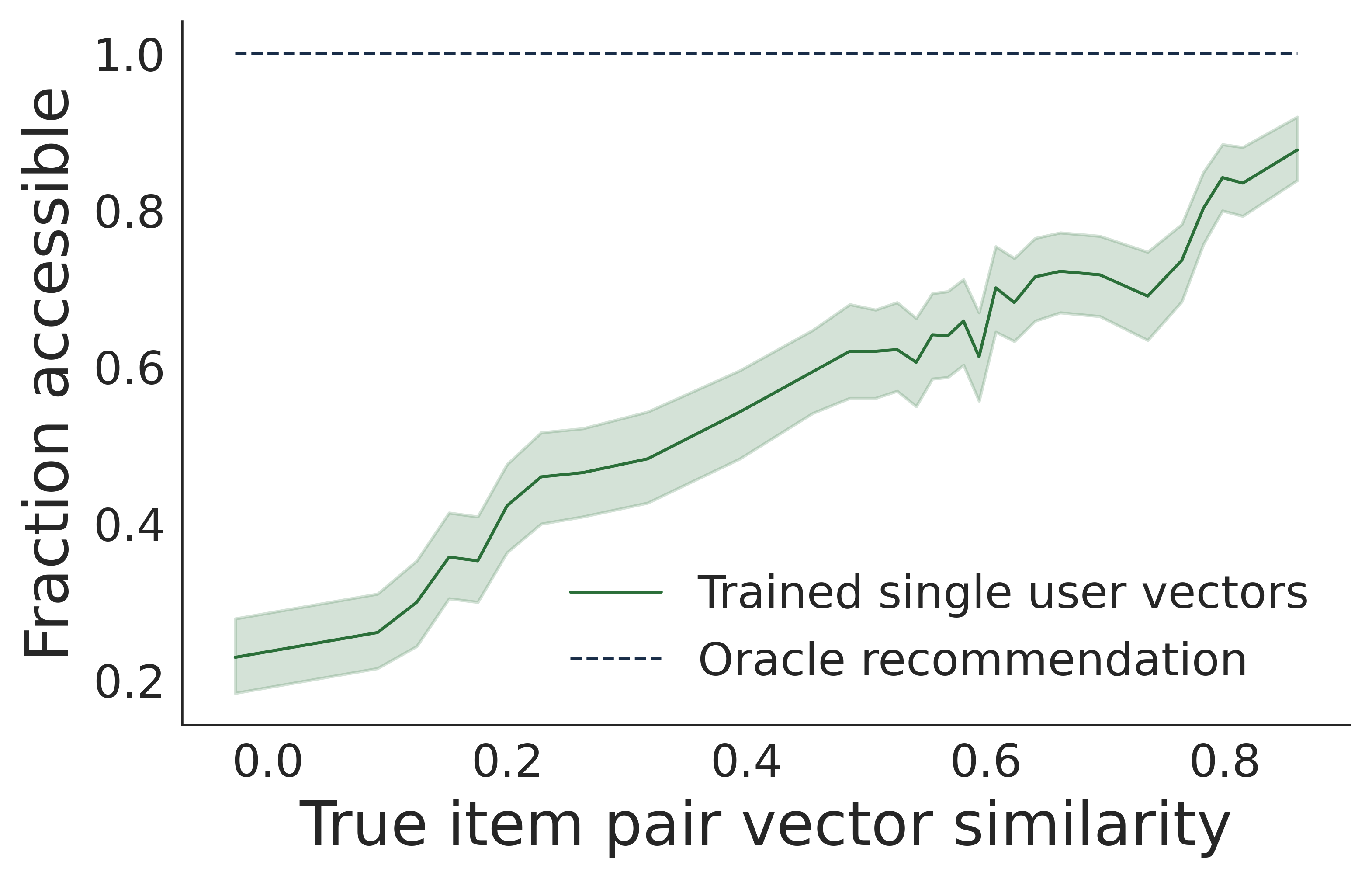}
		\label{fig:trainedimpossibilitysyn2}}
	
	\caption{Supplemental figures for the second synthetic experiment.}
	\label{fig:empsyn2min}  
	\Description{The left subfigure is a line graph with two lines, one for the oracle recommendations, and the other one for the trained single user vector recommendations. The y axis shows the average number of recommendations the item pairs with that similarity, ranging from 0 to 10, and the x axis shows the true item pair vector similarity, ranging from 0 to 0.8. The recommendation frequencies for both two lines overall increase with the item pair similarity, where there is a bump in the middle. The two lines overlapped a lot, showing that in this synthetic dataset, the trained system at this level is not recommending more similar pairs compared to the oracle recommendations. 
	The right subfigure is a line graph with two lines, one for the oracle recommendations, and the other one for the trained single user vector recommendations.  The y axis shows the fraction of accessible item pairs within all item pairs with a certain item pair similarity, starting from 0.2 to 1. The x axis shows the true item pair vector similarity, starting from 0 to 0.8. The fraction of accessible item pairs for the line for the trained single user vector recommendations increases with the item pair similarity, while the oracle line is always at 1. This shows that the accessibility of item pairs for the trained system is highly dependent on the item similarity, where more similar items are much more likely to be joint-accessible.}
	
\end{figure}

\end{document}